\documentclass[aps,amsmath,latexsym,amsfonts]{JHEP3}
\usepackage[dvips]{epsfig}
\usepackage{epsfig}
\usepackage[all]{xy}
\usepackage{graphicx}

\title{Matching conditions for a brane of arbitrary codimension}
\author{Christos Charmousis\\
LPT, Universit\'e de Paris-Sud, B\^at. 210, 91405 Orsay CEDEX, France}
\author{Robin Zegers\\ LPT, Universit\'e de Paris-Sud, B\^at. 210, 91405 Orsay
  CEDEX, France\\ APC, 11 place Marcelin Berthelot, F-75231 Paris Cedex 05
\footnote{UMR 7164 (CNRS, Universit\'e Paris 7, CEA, Observatoire de Paris)}}
\date{\today}
\abstract{We present matching conditions for distributional sources of arbitrary
codimension in the context of Lovelock gravity. Then we give examples,
treating maximally symmetric distributional $p$-branes, embedded in flat,
de Sitter and anti-de Sitter spacetime.
Unlike
Einstein theory,  distributional defects of locally smooth geometry and
 codimension greater than 2 are demonstrated to exist in Lovelock
theories.
The form of the matching conditions depends on the
parity of the brane codimension. For odd codimension, the matching conditions
involve discontinuities of Chern-Simons forms and are thus similar 
to junction conditions for hypersurfaces. 
For even codimension, the bulk Lovelock densities induce intrinsic Lovelock densities
on the brane. In particular, this results in the appearance of the induced
Einstein tensor for $p>2$. For the matching conditions we present, the effect
of the bulk 
is reduced to an overall 
topological solid angle defect which sets the Planck scale on the brane and to
extrinsic curvature terms.  
Moreover, for topological matching conditions and constant solid angle deficit, we find
that the equations of motion are obtained from an exact p+1
dimensional action, which reduces to an induced Lovelock theory for large
 codimension.  In essence, this signifies that the distributional part of the
 Lovelock bulk equations can naturally give rise to induced gravity terms on a
 brane of even co-dimension.
We relate our findings to recent results on codimension 2 branes.}
\maketitle
\preprint{hep-th/0502170\\
LPT-05-11}

\newcommand{\be}{\begin{equation}}
\newcommand{\ee}{\end{equation}}
\newcommand{\ba}{\begin{eqnarray}}
\newcommand{\ea}{\end{eqnarray}}
\newcommand{\bea}{\begin{eqnarray}}
\newcommand{\eea}{\end{eqnarray}}
\newcommand{\bean}{\begin{eqnarray*}}
\newcommand{\eean}{\end{eqnarray*}}
\newcommand{\bml}{\begin{mathletters}}
\newcommand{\eml}{\end{mathletters}}

\def\half{\textstyle{1\over2}}

\def\ie{{\it ie }}

\def\L{{\cal{L}}}

\def\k{{\cal{K}}}
\def\R{{\cal{R}}}
\def\C{{\cal{C}}}
\def\B{{\cal{B}}}

\def\pp{/ \hspace {-2pt} /}

\begin{document}

\section{Introduction}

Under what dimensionality and geometric conditions can we consider the 
embedding of distributional sources 
in a classical gravity theory? What are the resulting dynamics for
a brane sourced by its own energy-momentum tensor? 

In general relativity, the dynamics of a self-gravitating
hypersurface with a localised
distribution of matter is given by the junction conditions of
Israel \cite{israel} (see also \cite{darmois}). 
Continuity of the intrinsic metric and a
 jump in the extrinsic curvature, at the location of the
distributional source, account for the hypersuface's induced energy-momentum 
tensor. To first order, the dynamics of a thick domain wall is seen to
accurately obey 
the distributional equations \cite{bonjour1}.
For codimension
2 defects, \ie infinitesimal strings, the situation is a bit more
tricky. For distributional sources of
intrinsic maximal symmetry, for example straight cosmic strings,
it was 
found \cite{Vilenki} that their gravitational field is of topological 
character, \ie 
they induce
a locally flat geometry with a global angular defect  related to their
intrinsic tension. For curved infinitesimal strings, there
seemed to be a paradox summarised in the phrase: 
infinitesimally thin self-gravitating strings cannot bend (for discussions on
this, see \cite{israel2}). Geometrically, the worldsheet of a self-gravitating
string is totally geodesic rather than minimal. One way
round this paradox is to simply admit that the Dirac distribution description
ceases to be correct once the self-gravitating 
string is not maximally symmetric. If we want to include self-gravity effects, 
we can no longer neglect finite width corrections {\footnote{We thank
    B. Carter for discussions on this point.}}. 
In higher codimension, localised distributional defects are not permitted 
at all,
largely due to the presence of classical black hole theorems, which 
restrict the possible solutions and do not permit the embedding of
distributional sources. Mathematically, as the codimension increases, so
does the dimensionality
of the distribution we want to embed{\footnote{Typically in flat space 
we would have  $\delta^{(n)}\sim {\delta(r)\over r^{n-1}}$.}} 
and, not surprisingly, it is increasingly difficult
for Einstein's equations to account for such a term. Physically,
increasing
the  codimension renders the internal structure of
the source more and more important. For example,  
strings of differing energy density and
pressure cannot be described by a distributional 
source{\footnote{Even
    for codimension 1 defects, 
things get more complicated as we reduce the
   {\it relative} symmetries of the metrics we want to match. 
For example, one cannot match a
Kerr metric with flat spacetime because the bulk metric is not continuous.}}. 

In higher than four spacetime dimensions, the situation does not change much if we
stay in the context of Einstein's theory. Here, the most common context of
distributional sources is that of brane universes, which are often 
idealised codimension 1 or codimension 2 defects of localised matter. 
Wall type defects have been
exhaustively studied and we can successfully describe perfect fluid type
sources and evaluate analytically their cosmology \cite{bin}. 
It is only when
one touches at spacetimes of lesser symmetry, such as those of localised black
holes or anisotropic cosmologies, that the field equations become 
non-analytical 
(see for example \cite{emparan}, \cite{dani}). 
For codimension 2 defects, the
situation is far more delicate and one does not even have a description of
brane cosmology for perfect fluid Dirac sources (see for 
example \cite{cline}). This is in complete analogy to the case of cosmic
strings in 4 dimensions. The topological character of codimension 2 defects,
which has been related to self-tuning, raises
however some interesting questions regarding the cosmological constant problem
(see for example \cite{gariga}).

There is another option and that is to change the bulk gravity theory. We
then need at most second order field equations in order to have Dirac
sources with a continuous metric. The most general gravity theory which
is identical to Einstein's theory in 4 dimensions and satisfies 
the above property is Lovelock theory \cite{lovelock}. 
This theory obeys
energy conservation via generalised Bianchi identities 
and has a well defined perturbation operator \cite{Zwei} which ensures that the
graviton has the same number of deegres of freedom as in Einstein theory. 
The most familiar version of Lovelock's theory is, in 5 or 6
dimensions, Einstein-Gauss-Bonnet gravity, which has received some 
attention recently. 
The Gauss-Bonnet Lagrangian  
is the Euler characteristic in 4
dimensions. It is only for dimensions strictly greater than 4 that it yields
non-trivial dynamics. The nice properties of Lovelock theory are of geometric
origin. Every even dimensional manifold, compact say for simplicity,
 has a generalised  Euler characteristic, which is a topological invariant,
related to a curvature density under the Chern-Gauss-Bonnet
theorem. The dimensional continuation of this curvature density 
is the term to be included in the Lovelock 
action in one higher
dimension. The Lovelock action is thus the sum of increasing powers of the
curvature operator. However, the
intimate relation of the curvature densities to topology gives field equations which are 
 only up to and linear in second derivatives of the metric. 
Moreover, the theory
is unique, \ie there are no other curvature invariants with the above properties.  

The catch is that the field equations
in component language 
are of growing difficulty once we go further than the Einstein-Hilbert term. 
Nevertheless, static Lovelock black holes are known \cite{myers2},
\cite{boulware} which are pretty similar to their Einstein versions and 
classical black hole theorems such as Birkhoff's theorem \cite{wiltshire} 
and its extension have been shown to be true \cite{fax1} in
Einstein-Gauss-Bonnet theory. On the other hand, the 
construction methods of Einstein theory,
leading for example to the black
string, cannot be extended to Lovelock theory 
\cite{maartens} (see however the
recent work of \cite{tanaka} and the work of Wheeler \cite{crak}).
Also it has been recently shown in the context of braneworlds 
that Lovelock terms can
yield classical 
instabilities to well defined Einstein solutions with negative tension
branes \cite{fax2}.

In this paper, we show that Lovelock gravity has a
property that Einstein gravity in higher dimensions {\it does not have} and
that is to permit localised sources of arbitrary codimension. It has already
been shown, recently,  that Einstein equations could be obtained on
codimension 2 defects, if we allowed for the presence of the Gauss-Bonnet term
in the bulk Lagrangian \cite{Greg} (see also \cite{KS}, \cite{Kofinas:2004ae}, \cite{Papantonopoulos:2005ma}). Also,
there have been several interesting attempts to higher codimension braneworlds, from brane
intersections \cite{Kaloper}, \cite{Navarro}. Here, using the language of
differential  forms, in
order to evade technical difficulties, we  treat the general problem. 
We
demonstrate that, allowing for all Lovelock densities in the bulk, we can
obtain matching conditions for arbitrary codimension distributional p-branes.  
Unlike Einstein gravity, the possibility
of localising Dirac distributions in this theory are far richer and more
varied. We also 
give examples of maximally symmetric $p$-branes with locally regular geometry. 
These matching conditions will 
generalise the matching or junction conditions already known in the
literature. Furthermore, we will show that, for even codimension, the resulting equations of motion
for the p-brane can in some cases  
be derived from a $p+1$ dimensional
action containing only  induced curvature  terms and in particular the
Einstein term.

The basic reason for the existence of 
higher codimension defects is that 
higher order Lovelock densities will be seen to
permit the embedding of higher dimensional distributions. Thus one can have,
say, a Dirac 1-brane embedded in 6 dimensional spacetime. The gravitational field is
very similar to that of a cosmic string in 4 dimensions. Spacetime is locally
flat. It is only topologically non-trivial with a solid angle
defect removed from the 4 sphere (like a solid section cut-off from an
orange). In Einstein theory, cosmic p-brane solutions \cite{ruth} describe
p-branes in a spherically symmetric vacuum. They were found to be singular and
have been argued to describe a higher codimension defect. 
It is quite interesting that
a more complex theory such as Lovelock theory now yields simpler and locally
smooth solutions!

Mathematically, the derivation of the matching conditions follows Chern's proof
of the Gauss-Bonnet theorem for the normal geometry of the brane \cite{Chern}. The
basic idea is the following: in order to embed a distributional brane,  one
needs to integrate out the normal degrees of freedom. 
Since a Dirac distribution is insensitive to the
internal structure of the defect, one has to single out the 
terms which are invariant under changes of the geometry in the normal section. 
The locally exact
forms over the normal section will turn out to play the central role in
determining the matching conditions.

We will generically find that the matching conditions become less and less
restrictive as the codimension increases, which is not so surprising. 
Indeed, for a codimension 1 defect, one can
construct an integrable submanifold, \ie we can, due to Cauchy's theorem,
reconstruct the whole of spacetime at least locally, given the 
position of the surface and a well defined normal vector. In codimension 2 and
higher, this is no longer true. The matching conditions are simply {\it necessary}
conditions for the geometry of the defect but in no way are they {\it sufficient} to
reconstruct the
whole dynamics of the bulk spacetime. For instance, the matching conditions for
a codimension 2 braneworld are independent of the presence of a bulk
cosmological constant -- something which is related to the self-tuning properties
of codimension 2 defects.

In section 2, we give the generalised Gauss-Codazzi equations in form and
component formalism
and in section 3 we discuss the bulk dynamics in Lovelock gravity along with
Chern's extension of the Gauss-Bonnet theorem. In Section 4, we lay out the
mathematical tools and prove the matching conditions in form formalism. This
section can be skipped by readers less interested in the mathematical
proof and formalism. In section 5, 
we apply the matching conditions to already known cases and
develop new ones, giving in particular the induced actions and equations of
motion, where possible in component language. 
In Section 6, we give simple examples of matching
conditions in Lovelock gravity. We summarize our results and conclude. In the
appendix we include some additional mathematical definitions, proof of intermediate results
and exact form calculations to be used in the fourth section. 

\section{Generalised Gauss-Codazzi equations}
\label{codazzi}

In this section, we set our notations and give a self contained review of the
geometric quantities we will be using for the rest of the paper.
At the end of the section, we derive the generalised Gauss-Codazzi equations.

Let $(M,g)$ denote a smooth $D$-dimensional spacetime {\footnote{Throughout this
    section we suppose that the metric tensor components are $C^2$ functions
    locally. We will be relaxing this assumption in Section 4}} endowed with a metric $g$ 
and let $\Sigma$ be a timelike submanifold of dimension $p+1$ and codimension
$N$. In other words, $D=N+p+1$. Note that if $N=1$, $\Sigma$ is a timelike
hypersurface. To every point $P$ of spacetime $M$, we
associate an orthonormal local basis of the tangent space $T_P M$,
$(e_A)_{A \in [1,D]}$, such that
\be
g(e_A,e_B)=\eta_{AB}.\ee
We suppose furthermore that this basis is adapted to the submanifold $\Sigma$. 
Hence, for all P in $\Sigma$, if $(e_{\mu})_{\mu \in [1,D-N]}$ denotes
the local orthonormal basis of the tangent space 
$T_P \Sigma$ and $(n_I)_{I \in [1,N]}$
that of the normal sections 
$(T_P \Sigma)^{\perp}$, then we have $e_A=(e_\mu,n_I)$, the $n_I$ being
the $n$ normal vectors to the 
brane{\footnote{Capital Latin indices from the begining of the alphabet 
will run over indices of the bulk $M$. 
Small greek letters will run over indices of $\Sigma$, whereas Latin
indices $I,J,...$ will run over indices of the normal sections.} }.
Thus,
\ba
g(e_{\mu},e_{\nu})&=& \eta_{\mu \nu} \\
g(e_{\mu},n_{J})&=& 0 \\
g(n_{I},n_{J})&=& \delta_{IJ}.
\ea 
This decomposition of the local basis on $\Sigma$ 
can be extended locally, 
by continuity to the bulk spacetime.

In order to set up the generalised Gauss-Codazzi equations, we need to
define projection
operators onto $T\Sigma$ and $(T\Sigma)^{\perp}$ which we will 
call respectively $\pi_{\pp}$
and $\pi_\perp$.  Let
$\theta^{\mu}$ denote the dual basis of $e_{\mu}$ and
$\theta^{I}$ that of $n_I$, so that
$\theta^{\mu}(e_\nu)=\delta_{\mu}^\nu$ and
$\theta^{I}(n_J)=\delta_{J}^I$. The 1-forms
$\theta^A = (\theta^{\mu},\theta^{I})$ span the
$D$-dimensional vector space of 1-forms $\Omega^{(1)}(TM)$ of $M$. 
For an $(s,r)$-type tensor of $T^{(s,r)}M$, the relevant projection 
operators are 
\ba
\pi_{\pp} : T^{(s,r)}M & \rightarrow & T^{(s,r)}\Sigma \nonumber \\
\pi_{\perp} : T^{(s,r)}M & \rightarrow & (T^{(s,r)}\Sigma)^{\perp} \, , \nonumber 
\ea
projecting an arbitrary tensor of spacetime $M$ onto the tangent and normal 
sections of $\Sigma$. If $H$ is such a tensor of $M$, we
 will denote the projections by $H_{\pp}$ and
$H_{\perp}$ respectively. For example, for a $(1,1)$ tensor, we have the 
decomposition
$$
H=H^A{}_B e_A\otimes \theta^B=
H^\mu{}_\nu e_\mu\otimes\theta^\nu
+H^\mu{}_J e_\mu\otimes\theta^J
+H^I{}_\nu n_I\otimes \theta^\nu
+H^I{}_J n_I\otimes\theta^J
$$
and, obviously, $H_{\pp}=H^\mu{}_\nu e_\mu\otimes\theta^\nu$ and 
$H_{\perp}=H^I{}_J n_I\otimes\tilde{\theta}^J$.
The spacetime metric, a (0,2) tensor, is by definition
$$
g=\eta_{\mu\nu} \theta^\mu \otimes \theta^{\nu}+ \delta_{IJ}
\theta^I \otimes \theta^J \, .
$$
Via the projection $\pi_{\pp}$, it gives rise to the first
fundamental form $h$ of $\Sigma$,
\be
h=\pi_{\pp}(g)=\eta_{\mu \nu} \theta^{\mu} \otimes \theta^{\nu}.
\ee

Given a torsionfree metric connection $\nabla$ on $(M,g)$, we can define the
spacetime connection coefficients
\be
\nabla_{e_A} e_B = \Gamma_{AB}^C e_C.
\label{conM}
\ee
On the other hand, expanding on the adapted basis, we get 
the Gauss-Weingarten relations
\ba
\nabla_{e_{\mu}} e_{\nu} = \Gamma_{\mu \nu}^{\lambda} e_{\lambda} -
K_{\mu \nu}^I n_I && \qquad \nabla_{e_{\mu}} n_I = K_{\mu I}^{\nu} e_{\nu}
- \omega_{\mu I}^{J} n_J
\label{GW1} \\
\nabla_{n_I} e_{\mu}= -\tilde{\omega}_{I \mu}^{\nu} e_{\nu} + \kappa_{I
\mu}^{J} n_J && \qquad \nabla_{n_I} n_J = -\kappa_{IJ}^{\mu} e_{\mu} +
{\cal C}_{IJ}^{K} n_K \label{GW2},
\ea
where indices are raised and lowered using $\eta_{\mu \nu}$ and
$\delta_{IJ}$. These relations are important for they 
define the geometric quantities describing the embedding of
$\Sigma$ in $M$. For example, the connection coefficients of $\Sigma$
and of any integral submanifold of $(T\Sigma)^{\perp}$ are respectively
\be
\Gamma_{\mu\nu}^\lambda=\theta^\lambda(\nabla_{e_{\mu}} e_{\nu})
\qquad {\rm and} \qquad
{\cal C}_{IJ}^{K} =\theta^K(\nabla_{n_{I}} n_{J}) = \Gamma_{IJ}^{K}.
\ee
The extrinsic curvature components of $\Sigma$,
\be
K_{\mu \nu}^I=-\theta^I(\nabla_{e_{\mu}} e_{\nu}) =
- \Gamma_{\mu \nu}^I \, ,
\ee
symmetric in their greek indices,
are the components of the second fundamental form of $\Sigma$. Notice
that since we also have $K_{\mu I}^\nu =
\theta^\nu(\nabla_{e_\mu} n_I)$, the extrinsic curvature
measures the variation of the normal vectors along the tangential
directions of $\Sigma$, thus describing the curving of $\Sigma$ in $M$. The extrinsic twist components,
$$
\omega_{\mu I}^J=-\theta^J(\nabla_{e_{\mu}} n_I) = - \Gamma_{\mu I}^J \, ,
$$
are antisymmetric in their latin indices and vanish identically for
codimension 1 embeddings. The extrinsic twist describes how the normal
vectors are {\it twisted} around 
when we displace them in a tangent direction along
$\Sigma$. The same definitions can be made, with respect to $(T\Sigma)^\perp$,
for the respective twist $\tilde{\omega}_{I \mu}^{\nu}$ and extrinsic curvature
$\kappa_{IJ}^{\mu}$ components.
The induced connections on $\Sigma$ and on any integral manifold
of $(T\Sigma)^{\perp}$ are respectively defined using the spacetime connection 
$\nabla$ via the projectors,
\ba
\label{icpar}
\nabla_{\pp} &\equiv & \pi_{\pp}(\nabla) \\
\label{icperp}
\nabla_{\perp} &\equiv & \pi_{\perp}(\nabla) \, .
\ea
Both connections are, by construction, metric with respect to $h$ and
$g-h$.

A specification of the extrinsic and intrinsic geometries, satisfying
the Gauss-Weingarten equations, is generally not enough to define a
submanifold $\Sigma$. What we still miss are {\it integrability
conditions} relating the above geometric quantities to the ambient
curvature of the whole spacetime $M$. For the case of hypersurfaces,
these integrability conditions are related to
the Gauss-Codazzi equations. In order to obtain the generalised
integrability conditions, we first define the connection 1-form
and a curvature 2-form in
$M$, find their relevant projected counterparts on $\Sigma$ 
and finally relate them to the
ambient 
curvature 2-form.

The connection 1-form of $M$, $\omega^A{}_{B}$, is defined by
\ba
\omega^A{}_{B} &=& \Gamma_{CB}^A \theta^C
\ea
where $\Gamma_{CB}^A$ are the usual Christoffel symbols (\ref{conM}).
We can decompose $\omega^A{}_{B}$ in the tangent, mixed and normal
sections,
\be
\label{1form}
\omega^A{}_{B}= \left (\begin{array}{ccc}
\psi^{\mu}{}_{\nu} & \k^{\mu}{}_{I} \\
-\k^{I}{}_{\mu}  & \psi^I{}_{J}
\end{array} \right ).
\ee  
On using the Gauss-Weingarten relations (\ref{GW1}) and (\ref{GW2}) 
the ``tangent'' and ``normal'' connection 1-forms on $M$ are
\ba
\label{zi}
\psi^{\mu}{}_{\nu}&=&\Gamma^{\mu}_{\rho \nu} \theta^{\rho} -
\tilde{\omega}_{I \nu}^{\mu} \theta^I \nonumber \\
\psi^I{}_{J}&=&-\omega_{\mu J}^I  \theta^{\mu} + {\cal C}_{LJ}^I \theta^L.
\ea
Notice that both connection 1-forms are ``corrected'' by a term
due to the twist, not present for hypersurfaces. These are the connection
forms  associated to the local rotational invariance under $SO(N)$ and
$SO(1,D-N-1)$ of the local normal and tangent frames. Notice now that the
parallel projection $\psi^{\mu}_{\pp\nu}=\Gamma^{\mu}_{\rho \nu} \theta^{\rho}$
is just the induced connection 1-form of $\Sigma$ as one would
expect. Similarly using (\ref{GW1}) and (\ref{GW2}) we obtain,
\be
\label{ext12}
\k^I{}_{\mu}= K_{\nu \mu}^I  \theta^{\nu} - \kappa_{J \mu}^I
\theta^J.
\ee
which takes care of the mixed components in (\ref{1form}) 
and we call extrinsic curvature 1-form. Indeed its 
tangent projection on $\Sigma$ is related to the usual extrinsic curvature,
\be
\label{ext121}
\k^I_{\pp\mu}= K_{\nu \mu}^I  \theta^{\nu}.
\ee

The ambient curvature 2-form is defined by
\be
\label{curv}
\R^A{}_{B}=\frac{1}{2}R^A{}_{BCD} \theta^C \wedge \theta^D \, ,
\ee
with respect to the spacetime Riemann tensor.
It can be written as the sum of three linearly independent 2-forms defined
respectively by the projections 
on the parallel $(T\Sigma )^2$, mixed $T\Sigma \times (T\Sigma )^{\perp}$
and normal sections $(T\Sigma)^{\perp \,\, 2}$;
\be
\label{proj}
\R^A{}_B=\R^A_{\pp B} + \tilde{\R}^A{}_B + \R^A_{\perp B},
\ee
where
\ba
\R^A_{\pp B} &=& \frac{1}{2} R^A{}_{B \mu \nu} \theta^{\mu} \wedge
\theta^{\nu} \\
\tilde{\R}^A{}_B &=& R^A{}_{B \mu I} \theta^{\mu} \wedge
\theta^{I} \\
\R^A_{\perp B} &=& \frac{1}{2} R^A{}_{B IJ} \theta^{I} \wedge
\theta^{J}.
\ea
This decomposition will be essential in the next sections. On the other
hand, the spacetime curvature 2-form is linked to the connection 
1-form via the second Cartan structure equation
\be
\label{cartan}
\R^A{}_{B}=d\omega^A{}_{ B} + \omega^A{}_{C} \wedge 
\omega^C{}_{ B} \, .
\ee
In terms of the projections (\ref{1form}), this decomposes into the relations,
\ba
\label{gcmunu}
\R^{\mu}{}_{\nu} &=& d\psi^{\mu}{}_{\nu} + \psi^{\mu}{}_{\lambda} \wedge
\psi^{\lambda}{}_{\nu} - \k^{\mu}{}_I \wedge \k^I{}_{\nu} \\
\label{gcmui}
\R^{\mu}{}_I &=& d\k^{\mu}{}_I + \psi^{\mu}{}_{\nu} \wedge
\k^{\nu}{}_I + \k^{\mu}{}_J \wedge \psi^J{}_I\\
\label{gcij}
\R^I{}_J &=& d\psi^{I}{}_{J} + \psi^{I}{}_{K} \wedge
\psi^{K}{}_{J} - \k^{I}{}_{\mu} \wedge \k^{\mu}{}_{J}
\ea
These equations when projected into the parallel and perpendicular
directions yield the Gauss, Codazzi-Mainardi and Ricci-Voss equations, in
form formalism (see also \cite{Carter}, \cite{Capovilla}). 
Indeed, notice that upon parallel and perpendicular projection, the second
Cartan equation (\ref{cartan}) defines
\ba
\label{ic1}
\Omega^{\mu}_{\pp \nu} &=&\pi_{\pp}\left( d\psi^{\mu}{}_{\nu} + \psi^{\mu}{}_{\lambda} \wedge
\psi^{\lambda}{}_{\nu} \right ) \\
\label{ic2}
\Omega^{I}_{\perp J} &=& \pi_{\perp} \left (d\psi^{I}{}_{J} + \psi^{I}{}_{K} \wedge
\psi^{K}{}_{J} \right ) \, ,
\ea
where $\Omega_{\pp}$ and $\Omega_{\perp}$ stand for 
the induced curvature 2-forms respectively associated to the
induced metric connections (\ref{icpar}) and (\ref{icperp}). Therefore, the
Gauss equation, for example, is the parallel projection of (\ref{gcmunu}) and,
in form notation, it reads
\be
\label{gform}
\R^{\mu}_{{\pp}\nu}=\Omega^{\mu}_{\pp \nu}- \k^{\mu}_{{\pp}I} \wedge
\k^I_{{\pp}\nu} \, .
\ee
In the end,
using the parallel projection of (\ref{ext12}) and the definition (\ref{curv})
in (\ref{gcmunu}), we get the Gauss equation written in component form, 
\be
\label{G1}
R^\mu_{\pp\nu\lambda \rho}=R^{\mu(ind)}_{\pp \nu\lambda \rho}
-K^I_{\nu\lambda}K_{I\rho}^\mu+K^I_{\nu\rho}K_{I\lambda}^\mu \, ,
\ee
since by definition
\be
\label{led}
\Omega^{\mu}_{\pp \nu}=\half R^{\mu(ind)}_{\pp \nu\lambda \rho}\theta^{\lambda}
\wedge \theta^\rho \, .
\ee
Proceeding in a similar way, one can obtain the Codazzi equation 
from (\ref{gcmui}) whereas the Voss equation is obtained by the parallel
projection of (\ref{gcij}),
\be
\label{G3}
R^I_{\pp J\mu \nu}= - \nabla_{e_\mu} \omega^I_{\nu J} + \nabla_{e_\nu} \omega^I_{\mu J}  + \omega^{I}_{\mu K} \omega^{K}_{\nu J} -\omega^{I}_{\nu K}\omega^{K}_{\mu J}-K^I_{\mu\lambda}K_{J\nu}^\mu
+K^J_{\mu\lambda}K_{I\nu}^\mu
\ee
Notice that (\ref{gcij}) is analogous to
(\ref{gcmunu}) but for the normal frame. The relevant Gauss equation
can be obtained by using $\pi_{\perp}$,
\be
\label{G2}
R^I_{\perp J K L}=R^{I(ind)}_{\perp J K L}
-\kappa^\mu_{J K}\kappa_{\mu L}^I+\kappa^\mu_{J L}
\kappa_{\mu K}^I
\ee
We will be making extensive use of Gauss's equations (\ref{gform}), (\ref{G1}) and (\ref{G2})
later on in order to pass from the ambient curvature to the induced and
extrinsic curvature components of $\Sigma$.

\section{Euler densities and Lovelock Gravity: the dynamics of spacetime}
\label{Lovelock}

Having laid down the tools for studying the geometry
of the submanifold $\Sigma$ embedded in some spacetime $M$, we now 
review the dynamics of the bulk spacetime. In this section we give 
the classical field equations of Lovelock gravity which generalise those of Einstein
gravity in more than 4 dimensions.  
What will be essential for what follows is the 
link of Lovelock densities 
to the generalised Euler densities \cite{lovelock} of even dimensional manifolds.  
The dynamics of the submanifold will be addressed in the following section.

The Lovelock action can be simply expressed in the
language of forms (see for example \cite{Zumino}, \cite{mad}, \cite{Myers}, \cite{DM}).  
This is simple enough, particularly neat and very instructive so we do from scratch here. 
As noted in the previous section, the 
dual forms $\theta^A$ form a basis of the vector space of 
1-forms $\Omega^{(1)}(TM)$. Their antisymmetric
products can be used in order to construct 
a basis of the higher order forms acting  on
$TM$: any $k$-form $\omega$ in $\Omega^{(k)}(TM)$, 
where $(0\leq k\leq D)$, can be written
\be
\omega=\omega_{A_1 \cdots A_k} \theta^{A_1} \wedge \cdots \wedge \theta^{A_k}
\ee
with $\omega_{A_1 \cdots A_k}$ some smooth function.   Now 
define a (D-k)-form,
\be
\theta^{\star}_{A_1 \cdots A_k}=\frac{1}{(D-k)!} \epsilon_{A_1 \cdots A_k
A_{k+1} \cdots A_D} \theta^{A_{k+1}} \wedge \cdots \wedge \theta^{A_D}
\ee
where $\epsilon_{A_1 \cdots \cdots A_D}$ is totally antisymmetric in its
$D$ indices and $\epsilon_{12\cdots D}=1$. This quantity is called the Hodge
dual of the basis $\theta^{A_1} \wedge \cdots \wedge \theta^{A_k}$ of
$\Omega^{(k)}(TM)$. It defines a dual basis of $D-k$ forms of 
$\Omega^{(D-k)}(TM)$.
We can therefore write the Hodge dual of any k-form as,
\ba
\label{hodge}
\star : \qquad \qquad \qquad \Omega^{(k)}(TM) \qquad \qquad &\rightarrow & \qquad \Omega^{(n-k)}(TM) \nonumber \\
\omega= \omega_{A_1 \cdots A_k} \theta^{A_1} \wedge \cdots \wedge
\theta^{A_k} \quad
&\rightarrow & \quad \star \omega = \omega^{A_1 \cdots A_k} \theta^{\star}_{A_1 \cdots A_k}
\ea
Obviously, the wedge product of any form with its dual is a $D$-form,
proportionnal to the volume element of spacetime, $\theta^\star$. A
useful identity is
\be
\label{ident}
\theta^B\wedge
\theta^{\star}_{A_1...A_k}=\delta^B_{A_k}\theta^{\star}_{A_1...A_{k-1}}
-\delta^B_{A_{k-1}}\theta^{\star}_{A_1...A_{k-2}A_k}+ \cdots + (-1)^{k-1}\delta^B_{A_{1}}\theta^{\star}_{A_2...A_k}
\, .
\ee

This formalism allows for a closed definition of the action for 
Lovelock gravity. The k-th Lovelock lagrangian density
${\cal L}_{(k)}$ is defined by,
\be
\label{lk}
{\cal L}_{(k)}=\R^{A_1 B_1}\wedge \dots\wedge \R^{A_k B_k} \wedge
\theta^{\star}_{A_1B_1 \dots A_kB_k}= \bigwedge_{i=1}^k \R^{A_i B_i}
\wedge \theta^{\star}_{A_1B_1 \dots A_kB_k}
\ee
Clearly,  ${\cal L}_{(0)}$ is the volume element, 
whereas aplying (\ref{curv}) and the identity (\ref{ident}) gives
\be
\label{ein1}
{\cal L}_{(1)}= \R^{A_1 B_1} \wedge \theta^{\star}_{A_1B_1}=R \, \theta^\star
\, ,
\ee
the Ricci scalar density and
$$
{\cal L}_{(2)}= \R^{A_1 B_1} \wedge  \R^{A_2 B_2} \wedge 
\theta^{\star}_{A_1B_1A_2B_2}=(R^{ABCD}R_{ABCD}-4R^{AB}R_{AB}+R^2)\, \theta^\star 
$$ 
is the Gauss-Bonnet density. Given definition (\ref{hodge}), ${\cal L}_{(k)}$
is obviously a $D$ form and can thus be integrated over spacetime $M$. 
Note that for $k>D/2$, (\ref{lk}) vanishes so that if $D=4$ say, then  
${\cal L}_{(0)}$, 
${\cal L}_{(1)}$ and ${\cal L}_{(2)}$ are the only terms
present in the action (although ${\cal L}_{(2)}$, 
as we will now see, turns out to be trivial). 
Indeed, if $D$ is even,
for $k=D/2$, the Lovelock density $\L_{D/2}$ is simply the Euler density
of $M$, \ie 
\be
\label{euld}
{\mathcal L}_{(D/2)}= \bigwedge_{i=1}^{(D/2)} \R^{A_i B_i} \,
\epsilon_{A_1B_1 \dots A_{D/2}B_{D/2}} \, ,
\ee
whose integral over any D-dimensional compact manifold $M$ yields,
according to the Gauss-Bonnet-Chern theorem \cite{Chern}, 
the Euler characteristic (see for example \cite{spivak}) of $M$,
\be
\chi \left [ M \right ]= \frac{1}{(4\pi)^{D/2} (D/2)!} \int_{M} {\cal
  L}_{(D/2)} \, .
\ee
Since ${\mathcal L}_{(D/2)}$ is a closed form, 
it is also locally exact from Poincar\'e 's lemma and therefore does
not yield any contribution to the field equations. In other words, it is a pure divergence
term. 
For example in $D=4$, ${\mathcal L}_{(2)}$ stands for the Gauss-Bonnet density whose
integral is the 4 dimensional Euler characteristic. 
For a
$D=2k$-dimensional compact manifold $M$, with boundary $\partial M$, 
this result generalizes to
\be
\label{eul}
\chi \left [ M_{D} \right ]= \frac{1}{(4\pi)^{D/2} (D/2)!} \left [\int_{M_{D}}
{\cal L}_{(D/2)} + \int_{\partial M_{D}} \Phi_{(D,D/2-1))} \right ] \, ,
\ee
where the $(D-1)$-form $\Phi_{(D,D/2-1)}$ reads 
\cite{spivak,Chern,EGH,Gilkey,MH},
\be
\label{phi1}
\Phi_{(D,D/2-1)}= \sum_{m=0}^{D/2-1} \frac{D \cdot (D-2) \cdots 2(m+1)}{1\cdot
3 \cdots (D-2m-1)} \epsilon_{\mu_1 \cdots \mu_{D-1}} \left (\bigwedge_{l=0}^m
\R^{\mu_{2l-1} \mu_{2l}} \right )
\wedge \left ( \bigwedge_{l=0}^{D-1} \k^{\mu_l}{}_{N} \right ) \, .  
\ee
This is one of the Chern-Simons $(D-1)$-forms $\Phi_{(D,m)}$, which will play
a central role in the next sections. 
The first non-trivial such term  occurs for
$D=2$. In this case, the 1-form $\Phi_{(2,0)}$ reduces, up to a numerical
factor, to the mean extrinsic curvature of the boundary $\partial M_{1}$. In
higher dimensions, this form can be extended to a $D-1$ form, 
\be
\label{zztop}
Z_{(1)} = 2 \k^{\mu}{}_N \wedge \theta^{\star {\pp}}_{\mu}= 2 K \, \mbox{Vol}_{\partial M_{D}}
\ee
for $k=1$, where we have used (\ref{ext12}) and (\ref{ident}) and $N$ here denotes the
unique normal direction.
This can be recognised as the integrand of the 
Gibbons-Hawking boundary term \cite{GH}. For the
Gauss-Bonnet case $k=2$, the corresponding Chern-Simons form can be extended 
to 
\be
\label{my}
Z_{(2)}= 4 \left (\R^{\mu \nu} \wedge \k^\rho{}_N + \frac{2}{3} \k^\mu{}_N \wedge \k^\nu{}_N \wedge \k^\rho{}_N \right ) \wedge
\theta^{\star {\pp}}_{\mu \nu \rho} \, , 
\ee  
which coincides with the Myers-boundary
term \cite{Myers}.

The integral of the sum, for $k<D/2$, of all Lovelock densities ${\cal
  L}_{(k)}$ is the most general classical action for $M$, yielding up to
second order field equations for the metric tensor. This is the Lovelock
action, 
\be
\label{lla}
S_D= \int_{M} \sum_{k=0}^{[(D-1)/2]} \alpha_k {\cal L}_{(k)} \, ,
\ee
where the brackets stand for the integer part. 
The first three terms of this sum
\be
S_D = \int_{M} \left ( \alpha_0 \theta^{\star} + \alpha_1 \R^{AB}\wedge
\theta^{\star}_{AB} + \alpha_2 \R^{AB} \wedge \R^{CD}
\wedge \theta^{\star}_{ABCD} + \cdots \right )
\ee
are respectively the cosmological constant, Einstein-Hilbert and
Gauss-Bonnet terms, yielding the general classical action in
$D=5, 6$ dimensions. Notice that, with respect to the Euler
characteristic, there is a clean-cut seperation between
even and odd dimensional spacetimes, which will emerge again later on in
the matching conditions. 
 
A variation of the action (\ref{lla}) including matter, 
with respect to the frame, gives the Lovelock equations
\be
\label{ll}
\sum_{k=0}^{[(D-1)/2]} \alpha_k {\cal E}_{(k) A} = -2T_{AB} \theta^{{\star} B}
\, ,
\ee
where ${\cal E}_{(k) A}$ is the k-th Lovelock (D-1)-form,
\be
\label{klov}
{\cal E}_{(k) A}= \bigwedge_{i=1}^k \R^{A_i B_i} \wedge 
\theta^{\star}_{AA_1B_1 \dots A_kB_k}
\ee
and we have chosen the normalisation according to 
$$
-\half \R^{A_1 B_1} \wedge \theta^{\star}_{CA_1B_1}=G^A_C \theta^{\star}_{A} \, ,
$$
so that the equations of motion read, in component formalism, $G_{AB} +\cdots
= T_{AB}$. 
 For completeness, we note that 
variation with respect
to the parallel frame $\theta^{\mu}$ of the boundary terms (\ref{zztop}) and
(\ref{my}) leads to junction conditions for hypersurfaces \cite{Davis}, 
\cite{Grav}.

\section{Matching conditions}
\label{gauss}

We consider a distributional source
of codimension $N$ 
on the RHS of (\ref{ll}) and investigate under what conditions it can be
matched by some distributional geometric charge provided by the LHS of
Lovelock  equation. 
The general idea, that will guideline our search for matching conditions, 
follows the derivation of Gauss's theorem in electromagnetism: we
integrate (\ref{ll}) over an arbitrary  $N$ dimensional normal section to the
brane, which we denote $\Sigma^{\perp}$. Given that the source term is a Dirac
distribution, 
the value of this integral must not depend on the choice of geometry of the 
normal section. This is similar to neglecting the internal structure 
of a local Abelian-Higgs vortex or neglecting the 
internal structure  of a scalar field domain wall when investigating its zero
thickness limit. 
Our aim here is to set-up and seek the terms in the Lovelock
$(D-1)$-forms (\ref{klov}) which are independent of the geometry
of the normal section and can thus provide the distributional charge for
a Dirac distribution. This will lead us to the self-gravitating equations of
motion for $\Sigma$.

Let us therefore assume that the bulk energy-momentum is confined on a $(D-N-1)$-brane
which lies on the submanifold $\Sigma$ of arbitrary codimension $N$. 
The stress-energy tensor of the brane $\Sigma$ reads
\be
\label{set}
T_{AB}= \left ( \begin{array}{cc}
S_{\mu \nu} & 0 \\
0&0
\end{array} \right ) \delta_{\Sigma} \, ,
\ee
where $\delta_{\Sigma}$ is the Dirac distribution on $\Sigma$ as defined in
the appendix \ref{op}. We now consider the local existence of  
a normal section $\Sigma_P^{\perp}$ at point $P$,
which is an open submanifold of $M$, satisfying the following important 
properties:
\begin{itemize}
\item
$\Sigma_P^{\perp} \cap \Sigma = \{ P \}$
\item
$T_P\Sigma_P^{\perp} = (T_P\Sigma)^{\perp}$
\item
$\forall Q \in \Sigma_P^{\perp} \qquad T_Q\Sigma_P^{\perp} =
\mbox{Vect}(n_I)_Q$
\item $\overline{\Sigma_P^{\perp}}$ is homeomorphic to the $N$-dimensional
compact disk ${\mathbb D}_N$ and $\partial \Sigma_P^{\perp}
= \overline{\Sigma_P^{\perp}} \backslash \Sigma_P^{\perp}$
\end{itemize}
The second and third points mean that the normal vectors $(n_I)$ are locally
extended to the
bulk from $P$ so as to span the tangent bundle of $\Sigma_P^{\perp}$.
The last point means that $\partial \Sigma_P^{\perp}$ is homeomorphic to the
$(N-1)$-dimensional unit sphere ${\mathbb S}_{N-1}$, \ie has the same
topological properties (in the sense that open neighborhoods of
 $\partial\Sigma_P^{\perp}$ are mapped in open neighborhoods of ${\mathbb S}_{N-1}$). 
This mapping will be important later on when characterising 
the topology of $\partial \Sigma_P^{\perp}$. As an example,
consider the case of an infinitesimal cosmic string. In this
case,  $\Sigma_P^{\perp}$ is simply the two dimensional cone, whereas $\partial
\Sigma_P^{\perp}$ is the reduced deficit circle. 

We will have to integrate the equations of motion (\ref{ll}) over
$\Sigma_P^{\perp}$. Doing so requires
the projection of (\ref{ll}) in such a way as to get $N$-forms on
$T\Sigma^\perp$. The $(D-1)$-forms (\ref{klov}) and the stress-energy 
form $T_{AB} \theta^{\star B}$ must therefore be projected onto
$\Omega^{(D-N-1)}(T\Sigma) \otimes \Omega^{(N)}(T\Sigma^\perp)$, so that
we have an $N$-form on $T\Sigma^\perp$, the parallel factor behaving as a
$0$-form, {\it ie} as a mere function, with respect to integration over $T\Sigma^\perp$. 
For this purpose, we construct in Appendix \ref{op} a projection operator
$i_N$ (\ref{projector}), out of
the interior product $i$ on differential forms (\ref{int}). 
Applying (\ref{projector}) on the RHS of (\ref{ll}) gives us an
$N$-form over $(T\Sigma)^{\perp}$ as required,
\be
i_N \left (T_{AB} \theta^{\star B}\right ) = T_{A\nu}
\theta^{\star \nu}_{{\pp}} \wedge \theta^{\star\perp} \, ,
\ee
where $\theta^{\star \nu}_{{\pp}}$ denotes the tangent Hodge dual of the
tangent basis of $\Omega^{(D-N-1)}(T\Sigma)$ and $\theta^{\star\perp}$ is the volume element of $T\Sigma^\perp$. Therefore integrating over an arbitrary  normal section $\Sigma_P^\perp$ according to (\ref{set}), we have
\be
\label{tei}
\int_{\Sigma_P^{\perp}} i_N \left (T_{AB} \theta^{\star B}\right ) =
\int_{\Sigma_P^{\perp}} \,  S_{A\nu} \theta^{\star
\nu}_{{\pp}} \wedge \delta_{\Sigma} \theta^{\star\perp} 
= S_{A\nu}(P) \theta^{\star
\nu}_{{\pp}} \, .
\ee
Here, we see rather trivially that the matter quantity (\ref{tei}) is purely topological, {\it ie} its value does not
depend on the geometry of $\Sigma_P^{\perp}$. The projected equations
(\ref{ll}) give
\be
\label{match0}
\sum_{k=0}^{[(D-1)/2]} \alpha_k   \int_{\Sigma_P^{\perp}}  \; i_N \left({\cal
    E}_{(k) A}\right) = - 2  S_{A\nu}(P) \theta^{\star
\nu}_{{\pp}}
\ee
for any section $\Sigma_P^{\perp}$.  

We now seek all the $N$-forms in $i_N \left({\cal E}_{(k) A}\right)$, whose integral over 
$\Sigma_P^{\perp}$ (\ref{match0}) is independent of the geometry of
$\Sigma_P^{\perp}$. We 
proceed in two steps:
\begin{itemize}
\item{vary the induced geometry of $\Sigma_P^{\perp}$ while keeping the boundary fixed.}
\item{Calculate the integral on the LHS of (\ref{match0}) while shrinking the
    boundary to $P$.}
\end{itemize}

The intrinsic geometry of $\overline{\Sigma_P^{\perp}}$ is 
 fully encoded in the way we extend the normal vectors 
$(n_I)$ to the bulk, {\it ie}
in the choice of the normal connection{\footnote{Remember that ${\cal
 C}_{LJ}^I$ are the Christoffel symbols for $T\Sigma^\perp$.}} $\psi^I_{\perp J}={\cal C}_{LJ}^I
\theta^L$ (\ref{zi}), as well as the specification of the  boundary $\partial
 \Sigma_P^{\perp}$.  In the presence of a Dirac distribution, we expect a
 removable singularity at $P$. The different choices of normal connection
 $\psi^I_{\perp J}$ correspond to different regularised geometries in the
 neighborhood of  $P$
 (see figure 2)
 and allow us to use the Gauss-Codazzi formalism of section 2. Note that the
 geometries in question do not need to be axially symmetric, their regularity suffices. We now set-up
 the coordinate system for the smoothed out geometry:
in order to perform the boundary integrals over $\partial
 \Sigma_P^\perp$, we adapt the normal basis 
to the boundary $\partial \Sigma_P^{\perp}$.
We do it in such a way that the
 radial vector, 
$n_N$, is normal to the tangent space
 $T(\partial \Sigma_P^{\perp})$ and outward
 pointing (see figure 1). This latter choice induces an  orientation on $\partial
 \Sigma_P^\perp$ which is canonical (resp. reversed) in odd (resp. even)
 codimension, with respect to the orientation of $(n_1, \cdots, n_N)$. This
 choice also gives rise to a coordinate singularity at the location of the
 brane not to be confused with the removable singularity at $P$ (which has
 been smoothed out). Note in particular that the vector $n_N$ is rather special and can
 only be extended as a globally non-vanishing vector field over
 $\Sigma_P^\perp\backslash \{P \}$.  With respect to the embedding of the
 hypersurface $\partial \Sigma_P^\perp$ into  $\overline{\Sigma_P^{\perp}}$,
 this is just the GN gauge for the normal vector $n_N$ to $\partial
 \Sigma_P^{\perp}$.  Then, we can  decompose the normal
 connection 
\be
\label{1formperp}
\psi^I_{\perp J}= \left (\begin{array}{ccc}
\psi^{\dot{I}}_{\perp \dot{J}} & \psi^{\dot{I}}_{\perp N} \\
\psi^N_{\perp \dot{J}}  & 0
\end{array} \right ),
\ee
with $\dot{I}=(1,\cdots,N-1)$, much like we decomposed the bulk
connection in (\ref{1form}). 
In this decomposition, $\psi^{\dot{I}}_{\perp N}=K^{\dot{I}}_{\dot{J} N}\theta^{\dot{J}}$ constitutes the
extrinsic curvature 1-form of $\partial \Sigma_P^{\perp}$ and its components
are therefore singular at $P$. This is simply because $\partial
\Sigma_P^{\perp}$ is, by construction, homeomorphic to the sphere 
${\mathbb S}_{N-1}(\epsilon)$, whose extrinsic curvature goes like
$1/\epsilon$ at $P$. We will assume that all the other
components of the connection 1-form are regular on $\Sigma$, so that the
tangent and normal induced geometries are well defined along the
brane. This, in particular, implies that all the normal forms vanish at
$P$, except $\psi^{\dot{I}}_{\perp N}$, since in turn $\theta^{\dot{J}}\sim
\epsilon$ at $P$. In simple words we have smoothed out the defect's core with
all possible normal geometries described by $\psi^I_{\perp J}$.

Having taken care of the technicalities concerning the coordinate system, we
now vary the normal connection over a given $\Sigma_P^\perp$.
To do so, we consider the continuous set of normal connections
\be
\label{coc}
\forall t \in [0,1] \qquad \psi^I_{\perp J} (t) = \psi^I_{\perp J} (0) +
t \xi^I{}_J
\ee
with $\xi^{IJ}=-\xi^{JI}$. We further assume that 
$(\xi^{IJ})_{\partial \Sigma_P^{\perp}}= (\xi^{IJ})_{P}=0$, since at this stage, the
boundary is held fixed.
The important technical result of appendix \ref{cohom} demonstrates that the only forms $\Xi_{(N,k) \mu}$,
appearing in the LHS of (\ref{match0}), whose integrals are invariant under
(\ref{coc}) are {\it locally exact} with respect to $d_\perp$.

For the second step, we apply Stoke's theorem over $\Sigma_P^{\perp}$,
noting however, that $\Xi_{(N,k) \mu}$ are not {\it globally} exact over
$\Sigma_P^\perp$ since the tangent vector fields over this section are
singular at $P$. By a standard geometric procedure however, (see for example
tome 5 of \cite{spivak}) they can be pulled
back to the associated sphere bundle where they become {\it globally}
exact. This is due to the fact that the  sphere bundle  
admits, by construction, nowhere vanishing sections. 
Let us briefly describe this here, the details can be found in \cite{spivak}. 
The sphere bundle over $\Sigma_P^\perp$ is a manifold ${\mathcal S}_{2N-1}$
which is locally homeomorphic to $\Sigma_P^\perp \times {\mathbb S}_{N-1}$,
that is, if $\Sigma_P^\perp$ is covered by neighbourhoods $U_i$, then
${\mathcal S}_{2N-1}$ is covered by neighbourhoods $U_i \times {\mathbb
  S}_{N-1} $. A set of transition functions, which we shall not discuss here,
fixes the way the different neighbourhoods are patched together to cover
${\mathcal S}_{2N-1}$. The sphere bundle admits, by definition, a smooth
projection map 
$\pi_0 : {\mathcal S}_{2N-1} \rightarrow \Sigma_P^\perp$.
To all points $Q \in
\Sigma_P^\perp$, we associate $\pi_0^{-1}(Q) = {\mathbb S}_{N-1}$, the sphere
fiber above point $Q$. 
Conversely any vector field $s: \Sigma_P^\perp \rightarrow {\mathbb
  S}_{N-1}$ defines a non-vanishing section of the sphere bundle. 
The forms $\Xi_{(N,k) \mu}$ can
therefore be pulled back via $\pi_0^*$ to the sphere bundle, where there
will always exist some potential $(N-1)$-form $\sigma_{(N,k) \mu}$ such that
\be
\label{xi}
\pi_0^* \left ( \Xi_{(N,k) \mu} \right ) = d_\perp \sigma_{(N,k) \mu} \, .
\ee  
Note, on the other hand, that $n_N :\Sigma_P^{\perp}\backslash \{P \}
\rightarrow {\mathbb S}_{N-1}$ defines a section of the sphere
bundle. Therefore $\pi_0 \circ n_N$ is the
identity map over $\Sigma_P^\perp\backslash \{P\}$ and we have
\ba
\int_{\Sigma_P^{\perp}} \Xi_{(N,k) \mu} &=&
\lim_{\epsilon\rightarrow 0} \int_{\Sigma_P^{\perp}\backslash
B(P,\epsilon)} n_N^* \left ( \pi_0^* \,  \Xi_{(N,k) \mu}
\right )\nonumber \\
&=&\lim_{\epsilon\rightarrow 0} \int_{n_N \left (\Sigma_P^{\perp}\backslash
B(P,\epsilon)\right )} d_\perp \sigma_{(N,k) \mu} \, ,
\ea
where $B(P,\epsilon)$  is the ball of $\Sigma_P^\perp$ of radius $\epsilon$,
centered on $P$ (figure 1).

Now, we can apply Stoke's
theorem on the image of this section, to get\footnote{As discussed in
  appendix \ref{op}, $d_\perp$ is not the differential operator of
  $\Sigma_P^\perp$ but merely the projection of $d$. This is the origin of the overall minus sign.},
\ba
\label{ex}
\int_{\Sigma_P^{\perp}} \Xi_{(N,k) \mu} &=& (-1)^{D-N-1} \left (
\int_{\partial\Sigma_P^{\perp}} n_N^* \, \sigma_{(N,k) \mu} -
\lim_{\epsilon\rightarrow 0} \int_{n_N \left (\partial
B(P,\epsilon)\right )} \sigma_{(N,k) \mu} \right ) \nonumber \\
&=& (-1)^{D-N-1} \left (\int_{\partial\Sigma_P^{\perp}} n_N^* \, \sigma_{(N,k)
    \mu} - \mbox{ind}_P \left (n_N \right ) \int_{\pi_0^{-1} \left ( P \right
    )} \sigma_{(N,k) \mu} \right ) \, ,
\ea
where, in the last step, we have made use of the smoothness of the space-time
manifold inside $B(P,\epsilon)$\footnote{Thanks to the invariance of these
  integrals under (\ref{coc}), we can always smooth out the interior of
  $B(P,\epsilon)$, interpolating thus between $P$ and $\partial \Sigma_p^\perp$.} and of the definition of the index
$\mbox{ind}_P \left (n_N \right )$ of the vector $n_N$ at
$P$. The latter is the integer coefficient that relates the integral
over the sphere $n_N(\partial B(P,\epsilon))$ and the integral over the fiber
$\pi_0^{-1}(P)={\mathbb S}_{N-1}$ above point P, and it only depends on the shape of $n_N$
in the neighborhood of $P$. Due to the Poincar\'e-Hopf theorem \cite{spivak},
$\mbox{ind}_P \left (n_N \right )= \chi \left [\overline{\Sigma_P^\perp} \right
]$. This basically means that, once $n_N$ is adapted to the boundary, it is
the topology of $\Sigma_P^\perp$ that fixes the shape of $n_N$ around $P$,
\cite{spivak}. 

Let us pause at this point to make the connection
with Chern's proof of the Gauss-Bonnet theorem \cite{Chern}.
Indeed, (\ref{eul}) follows from (\ref{ex}), where 
we set $\Xi_{(N,k)\mu}={\mathcal L}_{(D/2)}$, the Euler density (\ref{euld}), 
and we show that $\pi_0^*
\, {\mathcal L}_{(D/2)} = d \Phi_{(D/2,D/2-1)}$
\cite{spivak,Chern}. Chern's form $\Phi_{(D/2,D/2-1)}$ is
 precicely the boundary term (\ref{phi1}). 
In our generic case, since the normal
section is homeomorphic to the unit disk, we know that
$\mbox{ind}_P \left (n_N \right )=1$. 

In the sequel, the second term in the
RHS of (\ref{ex}) will be called the {\it charge} of $\sigma_{(N,k) \mu}$ at $P$,
\be
\label{defq}
q_{(N,k)\mu}(P) \equiv (-1)^{D-N-1} \int_{\pi_0^{-1} \left ( P \right )} \sigma_{(N,k) \mu} \, .
\ee
Stepping back, it is clear that if the charge is zero, (\ref{ex}) reduces to
the usual Stoke's theorem, the form (\ref{xi}) being globally exact.
It turns out (see
Appendix \ref{appb}) that
all charged forms are given in
terms of the normal induced Chern-Simons 
$(N-1)$-forms over the associated sphere bundle \cite{Chern},
\be
\label{C-Sforms}
\Pi_{(N,m)}^\perp \equiv \epsilon_{\dot{I}_1 \cdots
\dot{I}_{N-1}} \left (\bigwedge_{l=1}^{m} \Omega_{\perp}^{\dot{I}_{2l-1}
\dot{I}_{2l}} \right ) \wedge \left (\bigwedge_{l=2m+1}^{N-1}
\psi^{\dot{I}_l}_{\perp N} \right ) \, ,
\ee
\ie they come from the $k\geq N/2$ terms. Note that for $m=0$ in (\ref{C-Sforms}), we
have the maximal power of $\psi^{\dot{I}}_{\perp N}$.
Furthermore, it is useful to define the $(N-1)$-form{\footnote{Note in
    particular that for $N=D$ and $i=D/2-1$, (\ref{z}) 
gives the familiar boundary term of the Lagrangian density (\ref{phi1}).} 
\be
\label{z}
\Phi^\perp_{(N,i)} = \sum_{m=0}^{i} \frac{2(i+1) \cdots 2(m+1)}{(N-2i-1)
\cdots (N-2m-1)} \, \Pi^\perp_{(N,m)} \, .
\ee
The relevant potential terms (see Appendix (\ref{appb}) for their derivation)
are decomposed in terms of (\ref{C-Sforms}) and are 
given by
\be
\label{cheese}
\sigma_{(N,k) \mu} =\sum_{j=0}^{\mbox{\footnotesize min} \left(
    [N/2],k-N+[N/2] \right) } \B_{(k,j)} \; \sigma_{(N,k,j) \mu}^{\pp} \wedge
    \Phi^\perp_{(N,[N/2]-1-j)} 
- \B_{(k,0)} \left (N-2\left [ N/2 \right ] \right ) H_{(N,k)\mu} \, ,
\ee
where for branes of odd codimension, we have an extra term (\ref{z1}). We
  have set
$$
\B_{(k,j)}={(-1)^{D-N-1} \, 2^{N-2[N/2]+2j} k!\over {(N-2[N/2]+2j)! 
(k-N+[N/2]-j)! ([N/2]-j)!}}
$$
and the parallel forms
\be
\label{cake}
\sigma_{(N,k,j)\mu}^{\pp}=\left (\bigwedge_{l=1}^{k-N+[N/2]-j} \R^{\lambda_{l} \nu_{l}}_{{\pp}} \right ) \wedge
\left ( \bigwedge_{l=1}^{N-2[N/2]+2j} \k^{\rho_l}_{{\pp} N} \right ) \wedge
\theta^{\star {\pp}}_{\mu \lambda_l \nu_{l} \rho_{l} } \, ,
\ee
which are powers of $\R_{\pp}$ 
and $\k_{\pp}$ and behave as  mere functions with respect to integration 
over the normal sections. 
The potential forms (\ref{cheese}), as seen from the bulk
spacetime, are $D-2$-forms of 
$\Omega^{(D-N-1)}(T\Sigma) \otimes \Omega^{(N-1)}(T\Sigma^\perp)$.
Finally, and this is important for the explicit calculations, the Lovelock
rank $k$, 
the spacetime dimension $D$ and the codimension $N$
must obey the inequality
\be
\label{special}
N/2\leq k \leq \left[ {D-1} \over 2 \right] \, ,
\ee
which selects in particular the Lovelock bulk terms $\L_{(k)}$ that can carry
a Dirac charge.

Equation (\ref{cheese}) is the main mathematical result of this paper. It
enables us to give the matching condition for a brane of arbitrary
codimension. To get an intuitive idea of (\ref{cheese}), note that when $k$
takes its minimal value, $k=N/2$ (for $N$ even), the potential reduces to 
\be
\sigma_{(N,N/2) \mu} =(-1)^{D-1} \theta^{\star {\pp}}_\mu \wedge 
\Phi^\perp_{(N,N/2-1)} \, ,\ee
which, remember, is related to the the potential for the Euler density,
\be
\label{steer}
\pi^*_0 {\mathcal L}^\perp_{(N/2)} = d_\perp \Phi^\perp_{(N,N/2-1)} \, ,
\ee
of the normal section $\Sigma^\perp_P$.
On the other hand, for even $N$ and 
$k\geq N/2$, the first term in the sum, $j=0$, of (\ref{cheese}) has parallel part
\be
\label{marion}
\sigma_{(N,k,0)\mu}^{\pp}=\left (\bigwedge_{l=1}^{k-N/2} \R^{\lambda_{l}
    \nu_{l}}_{{\pp}} \right )  \wedge \theta^{\star {\pp}}_{\mu \lambda_1
    \nu_{1} \cdots \lambda_{k-N/2}
    \nu_{k-N/2} } \, , 
\ee
\ie a power of the parallely projected curvature 2-form, and the perpendicular
    part is again the potential for the Euler density (\ref{steer}). According
    to Gauss relation (\ref{gform}), the form $\R_{\pp}$ is related to the
    induced curvature 2-form of $\Sigma$, $\Omega_{\pp}$. Therefore, in
    (\ref{marion}) and thus in (\ref{cheese}), we are {\it a priori} going to
    pick up the induced Lovelock densities of $\Sigma$ (\ref{klov}). If $N$ is
    odd, we will have an odd power of $\k$ in (\ref{cake}). We will
    explicitely see this in a moment.

\FIGURE{
\includegraphics[width=12cm]{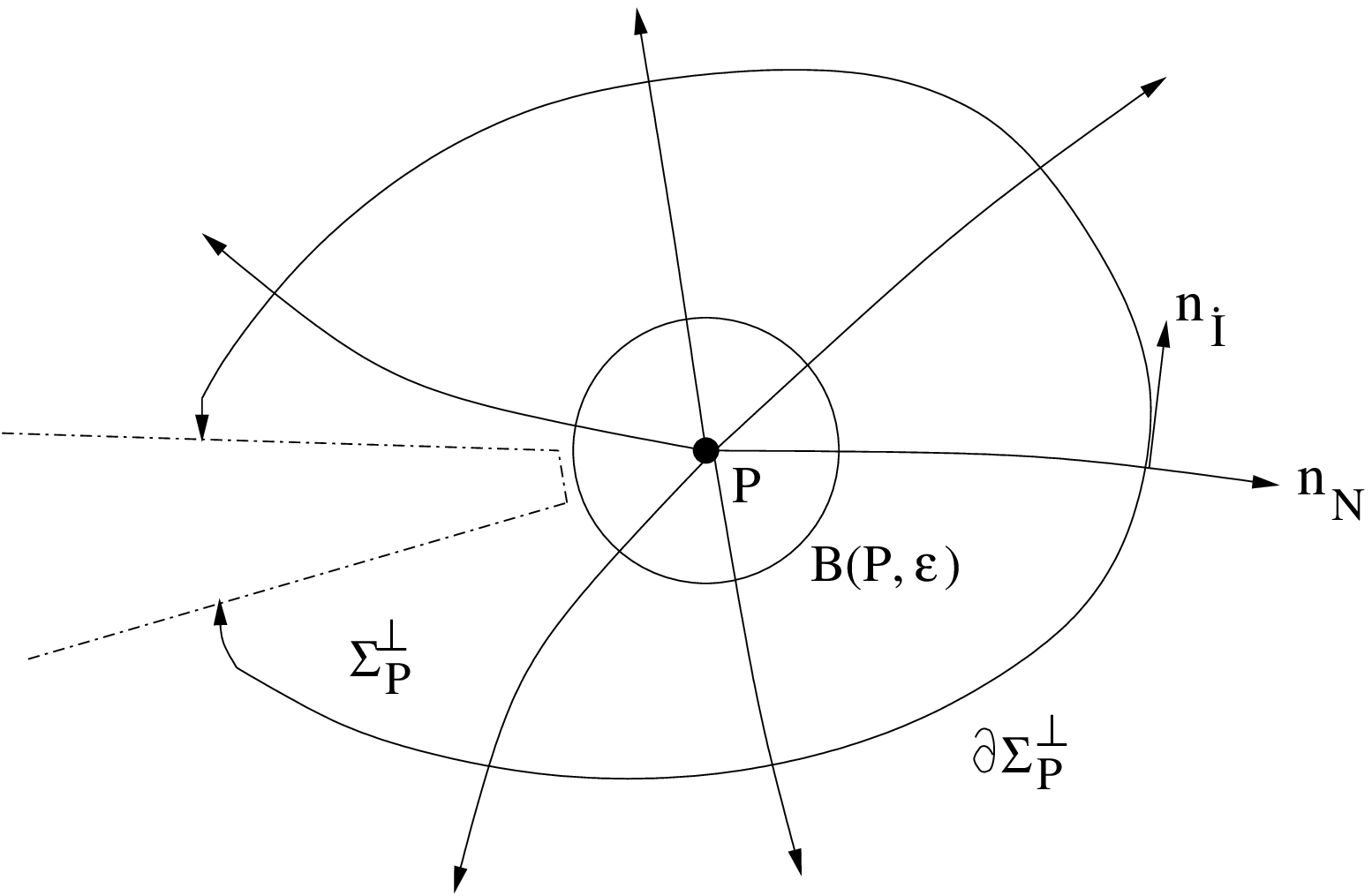}
\caption{Bird's eye view of the normal section $\Sigma^\perp_P$. The brane's
  internal dimensions are reduced to point P.}
\label{fig:min}}

We can now calculate the charge (\ref{defq}) at point
$P$. To do this,
 note that the integral is taken over the fiber ${\mathbb
  S}_{N-1}$ above $P$ and therefore the $(N-1)$-forms
(\ref{z}) will contribute through the maximal power 
of $\psi^{\dot{I}}_{\perp N}$, \ie through{\footnote{The sign originates
  from the orientation of the boundaries, which is canonical (resp. reversed)
  in odd (resp. even) codimension, so that $n_N$ is always outward pointing.}} 
$\Pi^{\perp}_{(N,0)} = (-1)^{N-1}(N-1)! \, \mbox{Vol}_{{\mathbb S}_{N-1}}$. We get, using (\ref{z}) and (\ref{cheese}),
\be
\label{residu}
q_{(N,k)\mu}(P) =(-1)^{N-1} \; D_{(N,k)}\;\mbox{Area}({\mathbb S}_{N-1}) \,
\sigma^{{\pp}}_{(N,k)\mu} (P) \, ,
\ee
where we have the parallel form
\be
\label{lemon}
\sigma^{{\pp}}_{(N,k)\mu}= \sum_{j=0}^{\mbox{\footnotesize min}\left
  ([N/2]-1,k-N+[N/2]\right )} \C_{(k,j)} \; \sigma_{(N,k,j) \mu}^{\pp}
\ee
and  
\be
D_{(N,k)}=\left\{ \begin{array}{cc} 2^{2n-1} k! (n-1)! & \mbox{if N=2n}\\ 
2^{n+3}k! & \mbox{{if N=2n+1}}
\end{array} \right .
\ee 
are constants independent of $j$. We also have
\be 
\label{mouche}
\C_{(k,j)}=\left\{ \begin{array}{cc} {1\over j! (k-n-j)!}  & \mbox{if N=2n}\\ 
{j! 2^{2j}\over {(k-n-j-1)! (2j+1)!}}  & \mbox{{if N=2n+1}}
\end{array} \right .
\ee
We show in Appendix C that the charge of $H_{(N,k)}$ is zero. 

\FIGURE{
\includegraphics[width=9cm]{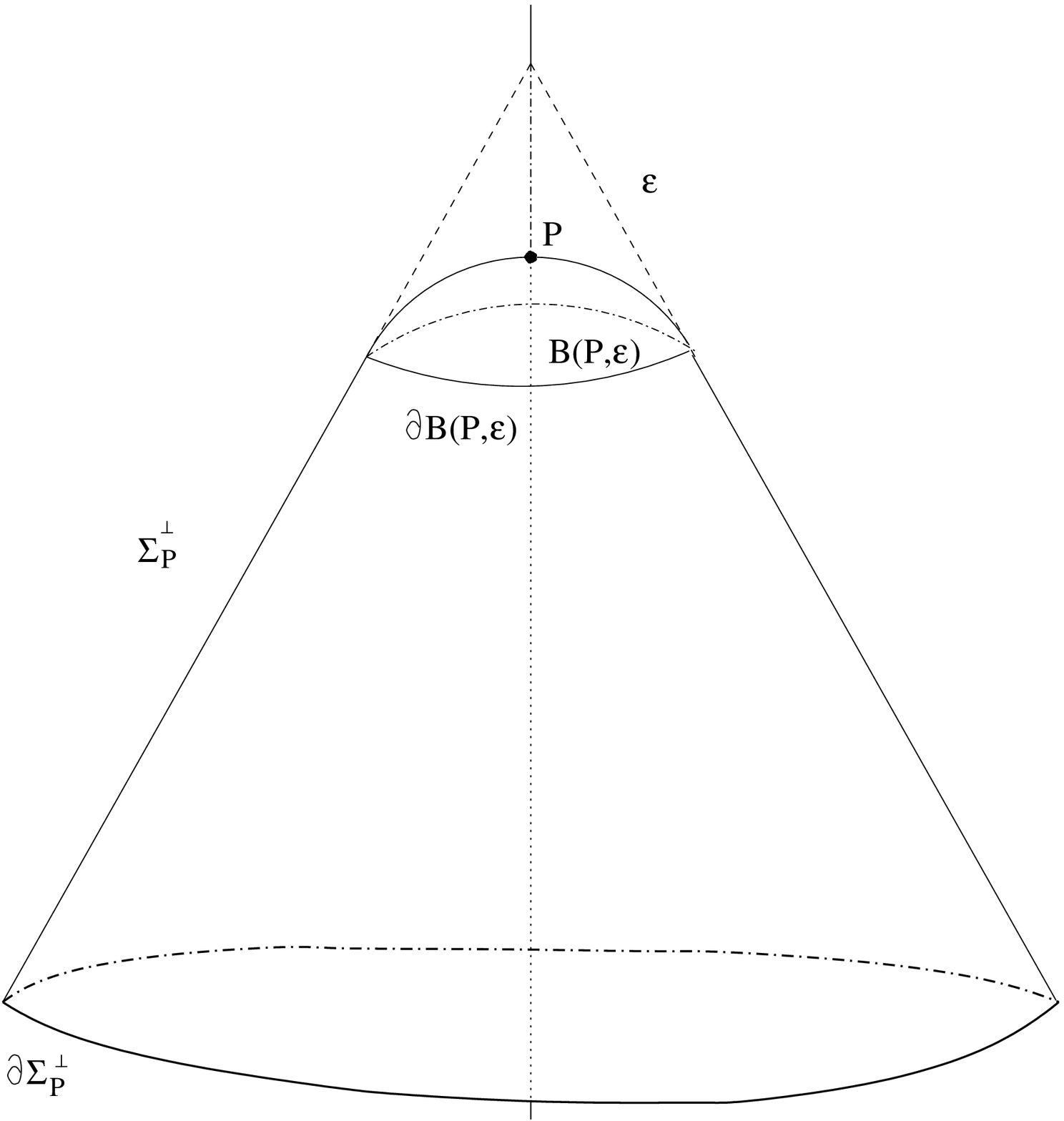}
\caption{The normal section $\Sigma^\perp_P$. The cone has been regularised
  here by an axially symmetric metric. Note that the first integral in
 (\ref{exdir}) refers to the boundary shrinking to the tip of the cone whereas
 the latter to the smoothed out charge at $P$. Their difference gives the
 geometric distributional charge.}
\label{fig:max}}
We can now write the matching conditions for a brane of arbitrary codimension,
\be
\label{matching}
\sum_{k=\left [ N/2 \right ]}^{\left [(D-1)/2\right ]} \alpha_k \left
[ q_{(N,k) \mu} \right ] (P) = - 2 S_\mu{}^\nu (P)
\theta^{\star {\pp}}_\nu \, ,
\ee
where
\be 
\label{exdir}
\left [ q_{(N,k) \mu} \right ] \equiv \lim_{\int_{\partial \Sigma_P^\perp}
 \rightarrow 0}  (-1)^{D-N-1} \int_{\partial \Sigma_P^\perp} n_N^* \, \sigma_{(N,k)\mu}
- q_{(N,k)\mu}(P) \neq 0 
\ee
is the discontinuity in the charge when passing from the
brane to the bulk and the potentials $\sigma_{(N,k)\mu}$ are given in (\ref{cheese}). 
The first and second terms in (\ref{exdir}) are of different nature if and
 only if the RHS of (\ref{matching}) is non-vanishing. In loose terms the
 former integral takes account of the Dirac singularity whereas the latter is the
 charge of the ``smoothed out defect''. For the case of the cosmic string, as
 we will see in the next section,
the former is the cone's exterior geometry giving the string defect $\beta$, whereas the latter is the
 charge of the smoothed out geometry giving simply 1 (see figure (\ref{fig:max})).
We remind the reader that $N$ is the codimension of the brane, $k$ is the rank
 of the Lovelock density (with $k=1$ for the Einstein term) and $D$ is the
 spacetime dimension.

In
the next section, we will put the matching conditions to application, 
starting with known cases and then focusing on cases of physical interest.
The important point to remember is that the potential (\ref{cheese}) is given
in terms of the Chern-Simons forms (\ref{C-Sforms}), which are integrated out
over the sphere, and a parallel part $\sigma^{{\pp}}_{(N,k)\mu}$, which will
end up giving the equations of motion for the brane.

\section{Action and equations of motion for a brane of arbitrary codimension}

Let us now analyse the matching conditions (\ref{matching}) at hand. In view
of the discontinuity (\ref{exdir}), we need to get a jump from the 
potentials $\sigma_{(N,k)\mu}$ as we shrink the boundary $\partial
\Sigma_P^\perp$ to $P$. Given the decomposition of $\sigma_{(N,k)\mu}$ into normal and parallel components (\ref{cheese}),
there are a priori three seperate ways to achieve this:
\begin{itemize}
\item{{\bf (I)} the parallel components $\sigma^{{\pp}}_{(N,k)\mu}$,
    (\ref{cake}), are smooth and the jump thus comes from the perpendicular
    part of $\sigma_{(N,k)\mu}$, \ie (\ref{z}). We refer to this case as the
    purely topological case for reasons that will become apparent in a moment.}
\item{{\bf (II)} the perpendicular part of $\sigma_{(N,k)\mu}$, \ie (\ref{z}),
    is smooth and the jump originates solely from $\sigma^{{\pp}}_{(N,k)\mu}$,
    (\ref{cake}).  This is always the case for matching conditions involving
    hypersurfaces, $N=1$, \cite{israel},\cite{Davis},\cite{Grav}}
\item{{\bf (III)} both parts in (\ref{cheese}) are discontinuous. For $N=2$
    and $D=6$, this case yields the results of \cite{Greg}.}
\end{itemize}

For (I), a discontinuity in the perpendicular
components originates from the Chern-Simons forms, (\ref{C-Sforms}). This
means, according to (\ref{residu}), that the area of the boundary surrounding the normal
section, \ie the limit of the integral in (\ref{exdir}), must be different
from the charge at $P$, which is proportional to $\mbox{Area}({\mathbb
  S}_{N-1})$, (\ref{residu}). Therefore, we must have a topological
defect at the location of $\Sigma$, permitting the discontinuity in (\ref{exdir}). For codimension 2, $N=2$, 
the topological defect is nothing but the conical deficit obtained by reducing the
circumference of ${\mathbb S}_{1}$ by $2\pi(1-\beta)$. Similarly, a topological
defect for any $N$ can simply be realised by cutting a solid angle $(1-\beta) \mbox{Area}({\mathbb S}_{N-1})$ 
from ${\mathbb S}_{N-1}$~\footnote{We shall see in the next sections precise examples of
this}, as one does for the {\it far} field of a global monopole, $N=3$, in Einstein
gravity (see for example \cite{Barriola}). For the topological case, using (\ref{exdir}), the matching conditions read 
\be
\label{topo}
(1- \beta(P) )\mbox{Area}({\mathbb S}_{N-1}) \sum_{k=\left [ N/2 \right ]}^{\left [(D-1)/2\right ]} (-1)^N \alpha_k
D_{(N,k)} \sigma^{{\pp}}_{(N,k)\mu}(P) = - 2  S_\mu{}^\nu (P)
\theta^{\star {\pp}}_\nu \, ,
\ee
where the equations of motion are essentially given by the geometric terms
(\ref{cake}) and $\beta$ {\it a priori} depends on the position $P$ along the
brane. This is clearly the simplest of cases and the one preserving the most
mathematical regularity for the spacetime metric, since both the induced and
extrinsic brane geometry are regular at $P$. We will argue later on in this
section that the
topological matching conditions are the ones adequate for even codimension
branes. 

In (II) on the contrary, the topology is trivial and the
discontinuity originates from the parallel components (\ref{cake}). In
particular, given that the equations of motion in the bulk are of second order
in the metric and that the metric is thus necessarily continuous at $P$, the
discontinuity can only originate from the extrinsic curvature 1-forms
(\ref{ext12}). This is
always the case with matching conditions involving hypersurfaces, $N=1$, since there is no geometry in the
normal sections, specifically $\Pi^\perp_{(1,0)}=1$.
The matching conditions read,
\be
\label{jump}
\mbox{Area}({\mathbb S}_{N-1})\sum_{k=\left [ N/2 \right ]}^{\left [(D-1)/2\right ]}  (-1)^N \alpha_k
 D_{(N,k)} \,
 \left( \lim_{\int_{\partial \Sigma_P^\perp}
 \rightarrow 0} \sigma^{{\pp}}_{(N,k)\mu|\partial\Sigma_P{^\perp}}
 -\sigma^{{\pp}}_{(N,k)\mu}(P) \right) 
= 2 S_\mu{}^\nu (P)
\theta^{\star {\pp}}_\nu \, .
\ee

In (III), the jump comes from a global discontinuity in (\ref{cheese}). Note
that this case is possible only because the perpendicular (\ref{z}) and the parallel components (\ref{cake}) in (\ref{cheese}) combine to
give a total differential obeying (\ref{xi}). This permits us to consider
distributions in the first place (see Appendix \ref{appb}). The importance of
this case will become manifest in the next
section, when treating specific examples for odd codimension defects.

As a first step, we now make the connection to known boundary conditions for Einstein and
Gauss-Bonnet 
gravity, starting with codimension $N=1$. 
The only normal induced Chern-Simons
form in (\ref{C-Sforms}) is the trivial one, $\Pi^\perp_{(1,0)}=1$, and, as we
mentioned above, a
charge can only originate from a discontinuity in the parallel part, case
(II). Replacing directly into (\ref{cheese}) and using (\ref{gform}) and
(\ref{ident}), we get in turn for the Einstein and Gauss-Bonnet terms,
\ba
\label{boss}
\sigma^{\pp}_{(1,1)\mu} &=& 2 \, \k^{\nu}_{{\pp}N } \wedge \theta^{\star
{\pp}}_{\mu \nu} = - 2 \, (K^{\nu}_{\mu } - \delta^{\nu}{}_\mu K ) \theta^{\star {\pp}}_{\nu} \nonumber \\
\sigma^{{\pp}}_{(1,2)\mu} &=& 4 \, \k^{\nu}_{{\pp}N} \wedge \left
( \Omega^{\rho \lambda}_{{\pp}} - \frac{1}{3} \k^{\rho}_{{\pp}N} \wedge \k^{\lambda}_{{\pp}N} \right ) \wedge \theta^{\star {\pp}}_{\mu \nu \rho
\lambda} \nonumber\\
&=& -4 (3 J_\mu^\nu-J \delta_\mu^\nu -2 P^{\nu}_{\;\;\lambda\rho\mu}
K^{\lambda \rho}) \theta^{\star {\pp}}_{\nu} \, ,
\ea
where we have set
$$
J_{\mu\nu} = \frac{1}{3}(2K K_{\mu\lambda}K^\lambda{}_\nu + K_{\lambda\rho}K^{\lambda\rho} K_{\mu\nu} 
{}- 2K_{\mu\lambda}K^{\lambda\rho}K_{\rho\nu} - K^2 K_{\mu\nu}),
$$ 
$$
P_{\mu\nu\lambda\rho} = R_{\mu\nu\lambda\rho} + 2 R_{\nu[\lambda} g_{\rho]\mu} - 2 R_{\mu[\lambda}
g_{\rho]\nu} + R g_{\mu[\lambda}g_{\rho]\nu}
$$
and we have dropped the label $N$ for the extrinsic curvature components.
Replacing into the matching conditions (\ref{jump}) and taking the
corresponding left and right limits (the charges (\ref{residu}) cancel out here), one gets the
well-known junction conditions for Einstein \cite{israel} and Einstein-Gauss-Bonnet gravity
\cite{Davis,Grav}. Not surprisingly, the equations of motion can be
derived from the boundary action, after variation with respect to the induced frame
$\theta^{\nu}$ on the left and right side of the brane. 
In differential form language the action is 
obtained trivially from (\ref{boss}),
\bea
S_{\Sigma}=2\alpha_1 \int_{\Sigma} \k^{\nu}_{{\pp} N} \wedge \theta^{\star{\pp}}_{\nu}
+4 \alpha_2 \int_{\Sigma} \k^{\nu}_{{\pp}N} \wedge \left
( \Omega^{\rho \lambda}_{{\pp}} - \frac{1}{3} \k^{\rho}_{{\pp}N} \wedge
\k^{\lambda}_{{\pp}N} \right ) \wedge \theta^{\star {\pp}}_{\nu \rho \lambda}
\eea
and agrees with the Gibbons-Hawking \cite{GH} and Myers \cite{Myers} boundary
terms (\ref{zztop}), (\ref{my}).

We now consider the codimension 2 case, $N=2$. Here, things are quite different for
the potentials of the charged exact forms (\ref{lemon}) coming from the Einstein and Gauss-Bonnet terms,
\ba
\label{einscodim2}
\sigma_{(2,1) \mu}^{\pp} &=& 2 \theta^{\star {\pp}}_\mu, \qquad \sigma_{(2,2) \mu}^{\pp} = 4 \R_{{\pp}}^{\nu \rho} \wedge \theta^{\star
{\pp}}_{\mu \nu \rho} \, . 
\ea
The charges are trivially obtained from (\ref{residu}) and they involve $\Pi^{\perp}_{(2,0)}$. This time,
the normal section has the topology of the unit 2-dimensional
disk. What is particularily interesting for the comprehension of the matching
conditions is that all three cases are possible for the
Gauss-Bonnet potential term. Indeed, 
starting with the Einstein term, we see from
(\ref{einscodim2}) that since the tangent frame is smooth,  
the associated charge (\ref{residu}) can only be purely topological,
\be
\left [q_{(2,1)\mu}\right ] = 4 \pi (1-\beta) \, \theta^{\star
{\pp}}_\mu \, ,
\ee
and refers to the conical deficit of $\Sigma_P^{\perp}$ -in particular, this
yields the matching
conditions for a cosmic string in $D=4$. For the
Gauss-Bonnet term, it is useful to decompose the projected curvature operator 
$\R_{{\pp}}^{\nu \rho}$ using Gauss relation (\ref{gform}),
\be
\label{GBcodim2}
\sigma^{{\pp}}_{(2,2) \mu} = 4( \Omega_{{\pp}}^{\nu \rho} \wedge \theta^{\star
{\pp}}_{\mu \nu \rho} - 
\k_{{\pp}}^{\nu I}\wedge \k_{{\pp}I}^{\rho} \wedge \theta^{\star
{\pp}}_{\mu \nu \rho}) \,  ,
\ee
where, note the summation over the normal index $I$.
Since the induced curvature 2-form $\Omega_{{\pp}}^{\nu \rho}$ of $\Sigma$
is smooth by hypothesis, the charge is topological
for the first term in the RHS of (\ref{GBcodim2}). For the second term
however, one can ask for a jump of the whole term, yielding the most
general contribution in (\ref{exdir}),
\ba
\label{banana}
\left [q_{(2,2) \mu}\right ] &=& 8 \pi (1-\beta) \, \Omega^{\nu
\rho}_{{\pp}} \theta^{\star {\pp}}_{\mu \nu \rho} + 8 \pi \, \left
[ \tilde{\beta} \, \k^{\lambda I}_{{\pp}} \wedge \k^{\rho}_{{\pp} I}\wedge
\theta^{\star {\pp}}_{\mu \lambda \rho} \right ]
\\
&=& - 16 \pi (1-\beta) G_\mu^{(ind) \nu} \theta^{\star {\pp}}_{\nu} + 16 \pi \, \left [ \tilde{\beta} \, W_{\mu}^{\nu} \right ]
\theta^{\star {\pp}}_{\nu} \, ,
\label{ananas}
\ea
where we have introduced the tangent induced Einstein tensor
$G_{\mu\nu}^{(ind)}$, and we have set \cite{Greg}, \cite{Kofinas:2004ae},
\be
\label{w}
W_{\mu}^{\lambda}=K^{\nu I}_{\mu }
K^{\lambda}_{\nu I} - K^{I} K^{\lambda}_{\mu I}  + \frac{1}{2}
\delta^{\lambda}{}_{\mu} \left( K_{I} K^{I} - K^{\nu I}_{\rho} K^{\rho}_{\nu I} \right )
\ee
 and 
\be
\tilde{\beta} \equiv  \left \{ \begin{array}{cc}
1 & \mbox{on the brane} \\
\beta & \mbox{in the bulk}
\end{array} \right .
\ee
This obviously corresponds to the 3rd bullet point.
The matching conditions (\ref{matching}) in this case read,
\be
\label{yann}
2 \pi \, (1-\beta(P))  \left \{ -\alpha_1 h_{\mu\nu}
 + 4 \alpha_2  G^{(ind)}{}_{\mu\nu} \right \} - 8 \pi \, \alpha_2
\, \left [\tilde{\beta}  W_{\mu\nu} \right ] = S_{\mu\nu}  \, .
\ee
This result generalises that derived in \cite{Greg} (see also
 \cite{Kofinas:2004ae}), where axial symmetry and the vanishing of the 
extrinsic curvature $K^\mu_{\nu N}$ on the
brane were assumed. These matching conditions ask for a singular behaviour of the extrinsic
 curvature terms, as well as a topological charge. This is rather constraining
 and mathematically not necessary. Let us now relax either of these assumptions.

If $\beta=1$, \ie there is no topological charge at all, the Einstein term
yields no contribution and the matching conditions are very much reminiscent
of those for hypersurfaces,
\be
-8\pi \, \alpha_2
\, \left [W_{\mu\nu} \right ] = S_{\mu\nu}  \, .
\ee
This is of course bullet point 1. Note that this case does not allow at all for matching
conditions for Einstein gravity  and can therefore be rejected as unphysical.
However, let us  suppose now that the extrinsic curvature terms are regular,
\ie $W_{\mu\nu}$ is continuous, and therefore the charge is only
topological. Note then that $W_{\mu\nu}$ on the brane only depends on the
normal extrinsic curvature, \ie we set $I=N$ in (\ref{w}), since, by
regularity, any angular
extrinsic curvature terms are zero on the brane's location. This 
yields the most natural case that agrees, on the one hand, with
the cosmic string calculation in Einstein gravity and, on the other hand, with maximal regularity of the
spacetime metric. We obtain,
\be
\label{tri}
2 \pi \, (1-\beta(P))  \left \{ -\alpha_1 h_{\mu\nu}
 + 4 \alpha_2  G_{\mu\nu}^{(ind)}  + 4 \, \alpha_2
\,  W_{\mu\nu} \right \} 
= S_{\mu\nu}(P)  \, .
\ee
Both for (\ref{yann}), \cite{Greg}, as well as for the topological matching condition
 (\ref{tri}), we see that 
 the bulk Einstein term, $k=1$, only allows for a pure tension brane, {\it ie} for a
cosmological constant term in the matching conditions, while the
Gauss-Bonnet term allows for arbitrary matter with an induced Einstein
equation 
on the brane \cite{Greg}. Some comments are now in order
 concerning the bulk degree of freedom $\beta=\beta(P)$. In particular, note that the matching conditions,
 \ie the distributional part of the Lovelock tensor,  do not permit us to
 evaluate this function. The remaining bulk equations have to be used in order
 to do so. Indeed, we emphasize that here, we have only calculated the distributional part of the Lovelock equations which
 yield the matching conditions. This, of course, is not sufficient to guarantee
 a full solution of the bulk field equations. Nevertheless, already at the level of (\ref{tri}), 
the effect of this scalar function is similar to a
 Brans-Dicke field, varying the coupling of gravity to matter on the brane and
 thus breaking the equivalence principle for a brane observer. Such effects are
 heavily constrained by local gravity experiments (see \cite{gdot}), where stringent
 observational bounds can be put on the variation in time of the function $\beta$. One also expects that, if $\beta$ is varying, the
 brane will be radiating gravitational waves into the bulk {\footnote{We
 thank Giorgos Kofinas for discussions on this.}}. These are very
 interesting issues which demand further work, but in order to present the matching conditions here, let
 us suppose for what follows that $\beta$ is constant.  In this case, there is an additional remarkable property
of the topological matching conditions, which is due to the fact that the extrinsic geometry is 
regular. The equations
of motion actually originate from a simple action involving induced and
 extrinsic quantities over $\Sigma$. In other
words, the degrees of freedom
associated to the normal section are completely integrated out, giving an exact
action for the brane's motion. In
differential form language, it is straightforward to read off the langrangian
 density in question. Literally, take out the free index for the charge in
(\ref{banana}) and use Gauss (\ref{G1}) to get,
\bea
\label{melina}
S_{\Sigma}^{(p=3,n=1)} &=& 4 \pi (1-\beta) \, \int_{\Sigma} \left( \alpha_1 \theta^{\star \pp}+ 2\alpha_2 (R^{\nu
\rho}_{{\pp}} \wedge \theta^{\star {\pp}}_{ \nu\rho} )\right)+\int_{\Sigma}
\L_{matter} \nonumber\\ &=& 4 \pi (1-\beta) \int_{\Sigma} \sqrt{-h} \; \left ( \alpha_1+ 2 \alpha_2
(R^{ind}-K^2+ K_{\mu\nu}^2) \right ) +\int_{\Sigma}
\L_{matter} \, ,
\eea
\ie the cosmological constant plus Einstein-Hilbert action with an extrinsic
curvature term.   
In other words, the only bulk quantity entering in the equations of motion is
the extrinsic curvature of the surface, giving a matter like component in the
action. If this is set to zero, \ie when $\Sigma$ is totally geodesic, we have
a cosmological constant plus Einstein gravity as an exact equation of motion
for the brane. Note that the overall mass scale of the integrated 4-dimensional theory is free and is set
by the topological defect $\beta$. Furthermore, we see a kind of pattern appearing between the Lovelock
densities in the bulk and the reduced ones on the brane. This ``reduction'',
from Gauss-Bonnet in the bulk ($k=2$) to Einstein
gravity on the brane 
and from Einstein in the bulk ($k=1$) to a cosmological term  on the brane, is a general feature that applies to
higher order Lovelock densities and higher even codimension. The origin is
again topological as we will see now.

Indeed, let us now assume that $N=2n$  is even. Let us also suppose that we have a
p-brane, \ie $D-2n=p+1$. This means, (\ref{special}), that the Lovelock rank
must satisfy
\be
\label{odd}
n\leq k \leq n+\left[ {p\over 2}\right ] \, .
\ee
For a 1-brane, we have $k=n$, \ie only a cosmological
  constant term $\tilde{k}=k-n=0$,
\be
S^{(p=1,n)}_{\Sigma}= (1-\beta) \mbox{Area}({\mathbb{S}}_{2n-1})) \,
\int_{\Sigma} \tilde{\alpha}_n \sqrt{-h} \, ,
\ee
where we have set $\tilde{\alpha}_k = D_{(2n,k)} \alpha_k$ and have taken
$\beta$ constant.
In turn, for a ($p=3$)-brane embedded in higher codimension, $n>1$, we have
$k=n$ and $k=n+1$. The topological matching condition (\ref{topo}) gives
\be
\label{3brane}
\tilde{\alpha}_n\theta_{\mu}^{\star\pp}+\tilde{\alpha}_{n+1}(\R_{{\pp}}^{\nu \rho} \wedge \theta^{\star
{\pp}}_{\mu \nu \rho} +  \k_{{\pp}N}^{\nu}\wedge \k_{{\pp}N}^{\rho} \wedge \theta^{\star
{\pp}}_{\mu \nu \rho})= \frac{2}{(\beta-1)\mbox{Area}({\mathbb S}_{N-1})} S_\mu{}^\nu (P)
\theta^{\star {\pp}}_\nu \, .
\ee
This equation tells us a remarkable property of 3-branes, for if we now use
Gauss equation (\ref{gform}), the ``squared'' extrinsic curvature form cancels out
leaving Einstein's equation with a cosmological constant!{\footnote{Note that
    for $n=1$, the ``squared'' extrinsic curvature  form in (\ref{3brane}) is
    absent.}} The 3-brane action is simply
\bea
\label{robin}
S_{\Sigma}^{(p=3,n>1)}&=& \mbox{Area}({\mathbb S}_{N-1}) (1-\beta) \int_{\Sigma} \sqrt{-h}\; \left (\tilde{\alpha}_n+ \tilde{\alpha}_{n+1}
R^{ind} \right )+\int_{\Sigma} \L_{matter}
\eea
and is valid as long as codimension is even and $n>1$. The Planck
scale{\footnote{If $\beta$ was varying its
variations would have to be small in order to satisfy local gravity
experiments \cite{gdot}}.} on the 3-brane is
\be
\label{stanley}
M_{Pl}^2=(1-\beta) \mbox{Area}({\mathbb S}_{N-1}) \tilde{\alpha}_{n+1} \, .
\ee
Note that only the topological matching conditions exist here, for there are no
extrinsic curvature terms at all! 
In a nutshell, we have a dimensional reduction from the Euler densities
of the bulk to the induced Euler densities on the brane. This has two
  important consequences. If we are for example in D=10 dimensions, it is the
highest order Lovelock density, $k=4$, that precicely gives us the Einstein
contribution on the brane. In turn, the $k=3$ density will give an induced
cosmological constant. The bulk Einstein term, for example, contributes nothing
since it cannot carry a Dirac charge! We
obtain the most general dynamics as far as the induced geometry of the brane
is concerned. The order of the induced Lovelock density is simply
  $\tilde{k}=k-n$ and obeys (\ref{odd})
\be
0\leq \tilde{k} \leq \left[ {p\over 2} \right] \, .
\ee

It turns out, \cite{z2}, that the extrinsic curvature terms identically drop
out, leaving a pure induced Lovelock brane action for any p-brane, as long as 
\be
\label{sofia}
\left[ {p\over 2}\right]+1\leq n \, .
\ee
This means that the worldvolume dimension is less than or equal to the brane's
codimension. If we now consider a 5-brane, we have $k=n$ which gives a
cosmological constant term ($\tilde{k}=0$), $k=n+1$ which gives the induced
Einstein term and no extrinsic curvature corrections ($\tilde{k}=1$), 
and in addition $k=n+2$ which yields a Gauss-Bonnet dynamical term on the
  brane ($\tilde{k}=2$). Explicitely, for $k=n+2$, we have the potentials
\bea
\sigma_{(2,3)\mu}^{\pp}&=&\R^{\nu \rho}_{{\pp}} \wedge \R^{\lambda
  \sigma}_{{\pp}}\wedge \theta^{\star {\pp}}_{\mu \nu\rho\lambda\sigma}\label{isa1}\\
\sigma_{(4,4)\mu}^{\pp}&=&\left( \half \R^{\nu \rho}_{{\pp}} \wedge \R^{\lambda
  \sigma}_{{\pp}}+  \R^{\nu \rho}_{{\pp}} \wedge \k^{\lambda}_{N
  \pp}\wedge\k^{\sigma}_{N\pp} \right) \wedge \theta^{\star {\pp}}_{
  \mu\nu\rho\lambda\sigma}\label{isa2} \, ,
\eea
which are valid for $n=1$ and $n=2$ respectively. Note then, that the first
term in each sum yields the Gauss-Bonnet tensor once we trade $\R^{\nu
  \rho}_{{\pp}}$ for the induced curvature 2-form $\Omega^{\nu \rho}_{{\pp}}$
by Gauss equation. For $n\geq 3$ (\ref{sofia}), the extrinsic
curvature terms drop out, giving us the induced Gauss-Bonnet tensor on the
brane,
\bea
\sigma_{(2n,n+2)\mu}^{\pp}&=&\half \left( \R^{\nu \rho}_{{\pp}} \wedge \R^{\lambda
  \sigma}_{{\pp}}+  2 \R^{\nu \rho}_{{\pp}} \wedge \k^{\lambda}_{N
  \pp}\wedge\k^{\sigma}_{N\pp}+ \right. \nonumber \\ 
&+&\left.  \k^{\nu}_{N
  \pp}\wedge\k^{\rho}_{N\pp}\wedge \k^{\lambda}_{N
  \pp}\wedge\k^{\sigma}_{N\pp} \right)\wedge \theta^{\star {\pp}}_{\mu
  \nu\rho\lambda\sigma}\nonumber\\
&=& \frac{1}{2} \Omega^{\nu \rho}_{{\pp}} \wedge \Omega^{\lambda
  \sigma}_{{\pp}}\wedge \theta^{\star {\pp}}_{\mu \nu\rho\lambda\sigma} \, .\label{isa3}
\eea  
Therefore, the 5-brane action reads
\bea
\label{5brane}
S^{(p=5,n\geq 3)}_{\Sigma} &=& \mbox{Area}({\mathbb S}_{N-1}) (1-\beta)
\int_{\Sigma} \sqrt{-h} \left ( \tilde{\alpha}_{n}+ \tilde{\alpha}_{n+1}
R^{ind} \right . \nonumber \\
&& \left . +\tilde{\alpha}_{n+2}
(R^{\mu\nu\lambda\rho}_{(ind)}R^{(ind)}_{\mu\nu\lambda\rho}-4R_{(ind)}^{\mu\nu}R^{(ind)}_{\mu\nu}+R_{(ind)}^2)
\right ) + \int_{\Sigma} \L_{matter} \, ,
\eea 
\ie the general Lovelock action for a 5-brane. This is a general feature for
even codimensions: {\it Any p-brane of even codimension, embedded in a $D\geq 2(p+1)-$dimensional spacetime,
obeys exactly the induced Lovelock equations!} At the level of zero thickness, there are
no degrees of freedom originating from the bulk spacetime apart from the
topological charge $\beta$. In particular, an
observer on a 3-brane will have no way of knowing that he is embedded in a 10
dimensional spacetime, up to finite width corrections.  Lastly, we argue
that in the case of even co-dimension one should consider topological matching
conditions. Indeed the first reason as we emphasized earlier on 
is mathematical regularity. There is no reason to impose discontinuous
geometric quantities if the Lovelock equations do not explicitely demand
it. The second reason comes from the nice physical properties of the
topological conditions. As we saw we can write down an induced action for
the brane which remarkably integrates out all the bulk degrees of
freedom. In particular, if condition (\ref{sofia}) 
is satisfied no extrinsic
curvature terms appear in the actual matching conditions and therefore only
topological matching conditions are possible.

For odd codimension, $N=2n+1$, we will concentrate only on the general matching
conditions (\ref{matching}). This is motivated by the maximally symmetric
example we will discuss in the next section. From (\ref{special}), we have
\be
\label{fax}
n+1\leq k \leq n+\left[ {p+1\over 2}\right] \, .
\ee 
We will always have odd powers of the extrinsic curvature form $\k$ in the
matching conditions, so there will never be an induced Einstein term on its
own. This is to be expected from the junction conditions for hypersurfaces.   
When $k=n+\left[ {p+1\over 2}\right]$, \ie $k$ has its maximal value, we have
$0\leq j \leq$min$(n,\left[{p+1\over 2}\right])$. Depending again on the
magnitude of $n$, we will pick up all or some of the geometric quantities. 
For the case of a 3-brane of codimension $N=3$, the relevant charges in (\ref{matching}) are
\bea
\label{fab}
q_{(2n+1,n+1)\mu}=\left[\tilde{\beta}
  \k^\rho_{N\pp} \right]\wedge \theta^{\star
  \pp}_{\mu\rho} \\
q_{(n=0,2)\mu}=q_{(n=1,3)\mu}=\left [  \tilde{\beta} \R^{\nu\sigma}_{\pp}\wedge 
  \k^\rho_{N\pp} \right]\wedge \theta^{\star \pp}_{\mu\nu\sigma \rho} \, ,
\eea
whereas if $N>3$, we obtain an additional term (\ref{fab}) for $k=n+2$,
  yielding 
\be
q_{(n>1,n+2)\mu}=\left ( \left [\tilde{\beta} \R^{\nu\sigma}_{\pp}\wedge 
  \k^\rho_{N\pp} \right]+ {2\over 3}\left[\tilde{\beta}\k^\nu_{N\pp}\wedge\k^\sigma_{N\pp}\wedge
  \k^\rho_{N\pp} \right] \right ) \wedge \theta^{\star \pp}_{\mu\nu\sigma
  \rho} \, ,
\ee
which is just an extension of the Myers boundary term \cite{Myers}, but for a
  3-brane. In other words, if 
\be
\label{simon}
\left[ {p+1\over 2 }\right]\leq n
\ee
is satisfied, then the matching conditions will only involve the dimensional
continuation of Chern-Simons forms on $\Sigma$.

\section{Maximally symmetric 1-branes of higher codimension}

The matching conditions we discussed in all generality in the previous section
are only necessary conditions for the existence of higher codimension
Dirac branes. Indeed, the matching conditions only depend on  the
$\mu\nu$-components of the Lovelock equations (\ref{ll}) and unlike
hypersurfaces do not give Cauchy type conditions for the local reconstruction
of the bulk spacetime. 
In this section, we construct simple spacetime solutions that carry a
brane of Dirac charge. We concentrate on cases of maximal symmetry both for
the bulk and for the brane. Such solutions, as we will argue, do not exist for
Einstein gravity.

\subsection{The codimension 3 case}

Consider a maximally symetric 1-brane embedded in a 5-dimensional static
spacetime of axial symmetry
about the brane. We call $r$, the proper distance to the brane. Then, let $A(r)$ be the brane warp factor and $L(r)$ the
size of the 2-spheres. We thus have
\be
\label{cod3a}
ds^2=A^2(r) \left ( -dt^2 + e^{2kt} \, dz^2 \right ) + dr^2 + L^2(r)
\left (\beta^{-2} \, d\theta^2 + \sin^2 \theta \, d\phi^2 \right ) \, .
\ee
Note that in order for the induced geometry of the brane to be adS, one can
Wick rotate $t$ and $z$ in (\ref{cod3a}), to obtain a flat slicing of adS.
The stress-energy tensor of the brane located at $r=0$, is taken to be
\be
\label{bull}
T^t{}_t = T^z{}_z = \sigma \, \frac{\beta \delta (r)}{4\pi L^2} \, ,
\ee
with all the other components vanishing.
We have odd codimension and, since $n=1$, we have Lovelock density up to
$k=2$. From (\ref{fax}), we know that the only possible charge comes from the
Gauss-Bonnet term $\L_{(2)}$, whose components read 
\ba
H^t{}_t = H^z{}_z &=& \frac{4}{AL^2} \hat{f}'+  \frac{4}{A} \left [f\right]
{\delta(r)\over L^2}\label{cyril}\\
H^r{}_r &=& \frac{4 L'^2 \left ( k^2-3A'^2 \right ) + 4 \beta^2 \, \left
    (A'^2- k^2 \right )}{A^2 L^2}\label{fox} \\
H^\theta{}_\theta = H^\phi{}_\phi &=& 4 \frac{\left ( \left (k^2-A'^2 \right
) L' \right )'}{A^2 L} \, ,\label{yolanda}
\ea
with $f=A' \left (\beta^2 -  L'^2 \right )$. As usual, the square brackets
denote the jump and $\hat{\mbox{ }}$ the smooth part. 
Given  (\ref{cyril}), we can get a Dirac distribution centered on the world-sheet, provided
that $f$ is discontinuous at $r=0$. {\footnote{Note that
(\ref{yolanda}) cannot carry a 3 dimensional distributional charge since it is
not divided by a sufficient power of $L$.}}  To make things
simple, let us first consider a pure Gauss-Bonnet equation,
$H^A_B=0$, since this is the only term that will provide a charge.  
The general solution reads,
\be
\label{solid}
L(r) = lr \, , \quad A(r)= a r+r_0 \, .
\ee
There are now at least two different ways, using the spacetime topology, to
obtain a distributional charge. We either introduce a uniform solid angle
defect (like the global monopole) for the entire spherical section or we
introduce a defect angle in the direction of the axis $\theta=0$. 
The former is achieved by setting $\beta=1$ in (\ref{cod3a})
and $l\neq 1$. However, the
solutions (\ref{solid}) generically have a curvature singularity{\footnote{Note here that
    finding smooth solutions to the pure Gauss-Bonnet equations (\ref{cyril}-\ref{yolanda}) does not
    guarantee the absence of singularity in the Riemann curvature!}} as $r\rightarrow 0$, as can be seen from the divergence of the Ricci scalar,
\be
R \sim -8 \,  \frac{ A'}{A} \frac{L'}{L} + 2 \, \frac{\left (\beta^2 - L'^2
  \right )}{L^2} \, .
\ee
In order to avoid such a singularity in the vicinity of the brane, we consider
the latter case,
\be
L' \sim \beta \quad \mbox{and} \quad A'\sim 0 \, ,
\ee
as $r$ goes to zero and therefore 
\be
\label{Lcone}
f(r) = \left \{ \begin{array}{cc}
0  & \mbox{if $r = 0^+$} \\
a (1-\beta^2) & \mbox{if $r=0$}
\end{array} \right . 
\ee
to get the desired Dirac distribution. Note here that the matching
conditions have to be mixed (III)! They cannot be only topological 
for then we would have a singular spacetime metric. Let us also point
out that the matching conditions do not suffice to guarantee the good
behaviour of the spacetime metric. Replacing directly into (\ref{match0}), we get in turn,
\be
\alpha_2 \int_{\Sigma_P} {dr\, d\theta\, d\phi L^2 \sin \theta \over \beta}
[A'(\beta^2-L'^2)]{\delta(r)\over A L^2}=-{4\pi \alpha_2 \over \beta
  r_0}a(\beta^2-1)={\sigma} \, ,
\ee
where the powers of $L$ drop out, giving the usual 1-dimensional Dirac
distribution. In the end,
\be
\label{spm}
4\pi \alpha_2 {a\over \beta r_0}(1-\beta^2)=\sigma 
\ee
and we have a pure Gauss-Bonnet 1-brane of codimension 3. We should note
however at this point that our construction introduces a 
conical singularity along the polar
axis $\theta=0$, not only at $r=0$.

Now that we have understood how
to build a Dirac distribution out of the Gauss-Bonnet term, we can
extend the simple calculation to the full Lovelock theory, with field equations
\be
\Lambda \delta^A{}_B + \zeta \, G^A{}_B + \alpha \, H^A{}_B = 0 \, .
\ee 
We get in the bulk,
\ba
\Lambda &+& \zeta \, \left (\frac{A''}{A} + 2 \frac{A'}{A} \frac{L'}{L} +
2\frac{L''}{L} + \frac{L'^2 - \beta^2}{L^2} \right ) + 4 \alpha \,
\frac{\left (A' \left (\beta^2- L'^2 \right ) \right )'}{AL^2}  = 0 \\
\Lambda &+& \zeta \, \left (\frac{A'^2 - k^2}{A^2} + \frac{L'^2 -
\beta^2}{L^2} + 4\frac{A'}{A}\frac{L'}{L} \right )+ 4\alpha \,
\frac{L'^2 \left (k^2- 3A'^2 \right ) + \beta^2 \, \left ( A'^2-k^2 \right )}{A^2 L^2} = 0 \\
\Lambda &+& \zeta \, \left (2 \frac{A''}{A} + \frac{L''}{L} + \frac{A'^2 -
k^2}{A^2}+2\frac{A'}{A}\frac{L'}{L} \right )+ 4 \alpha \, \frac{\left
( \left (k^2-A'^2\right ) L' \right )'}{A^2 L} = 0 
\ea
which admits the following solution,
\be
\label{ds}
ds^2 = k^2 r_0^2 \, \cos^2 \left (\frac{r}{r_0} \right ) \, \left ( -dt^2 +
e^{2kt} dz^2 \right ) + dr^2 + r_0^2 \beta^2 \, \sin^2 \left (\frac{r}{r_0}
\right ) \, \left ( \beta^{-2}d\theta^2 + \sin^2 \theta \, d\phi^2 \right ) \, ,
\ee
with
\be
r_0^2 = \frac{\zeta \mp \sqrt{\zeta^2 + \frac{4 \alpha
\Lambda}{3}}}{\Lambda / 3} \, .
\ee
This is a slicing of de-Sitter spacetime for $\Lambda>0$. The flat
space limit{\footnote{The constant $k$ can be gauged away in (\ref{ds}).}} is
obtained by taking $k\sim1/r_0$ and $r_0\rightarrow \infty$. Then, spacetime
is locally flat, exhibiting only a conical singularity at $r=0$, as for a
straight cosmic string \cite{vilenkin}. The negative cosmological constant case is simply
\be
\label{ads}
ds^2 = k^2 r_0^2 \, \cosh^2 \left (\frac{r}{r_0} \right ) \, \left ( dz^2 -
e^{2kz} dt^2 \right ) + dr^2 + r_0^2 \beta^2 \, \sinh^2 \left (\frac{r}{r_0}
\right ) \, \left ( \beta^{-2}d\theta^2 + \sin^2 \theta \, d\phi^2 \right ) \, ,
\ee
with
\be
r_0^2 = \frac{\zeta \mp \sqrt{\zeta^2 + \frac{4 \alpha
\Lambda}{3}}}{-\Lambda / 3} \, .
\ee

Now, we can use the same method as in the pure Gauss-Bonnet case
(\ref{Lcone}) to obtain matching conditions for distributional matter (\ref{bull}). 
Going through all the terms in the field equations, we see again that the only
term carrying a distributional charge is  $f$ (\ref{Lcone}). Terms originating
from the Einstein tensor, like ${L''\over L}$, could carry a 2 dimensional
Dirac charge but not a 3 dimensional one.  We have therefore constructed 
an Einstein-Gauss-Bonnet 1-brane of codimension 3, embedded in a
constant or zero curvature bulk. The matching conditions are again given by (\ref{spm}).

\subsection{The codimension 4 case}

We  now consider a 1-brane of codimension 4 as a solution of pure Gauss-Bonnet gravity. The 6 dimensional
space-time metric with axial symmetry reads,
\ba
\label{codim4metric}
ds^2 &=& A^2(r) \left ( -dt^2 + e^{2kt} \, dz^2 \right ) \nonumber \\
&&+ dr^2 + L^2(r)
\left (\beta^{-2} \, d\theta^2 + \gamma^{-2} \, \sin^2 \theta \, d\phi^2
+ \sin^2 \theta \, \sin^2 \phi \, d\psi^2 \right ) \, ,
\ea
and we assume the presence of an energy-momentum tensor,
\be
T^t{}_t = T^z{}_z = \sigma \frac{\beta \gamma \delta (r)}{2\pi^2L^3} \, , 
\ee 
As for the codimension 3 case, $L(r)\sim lr$ as $r$ goes to zero and the metric (\ref{codim4metric}) is
generically singular at $r=0$,
\be
R \sim -12 \, \frac{A'}{A} \, \frac{L'}{L} + 2 \, \frac{3 L'^2 -
2\beta^2 -\gamma^2 + 3 \left ( \beta^2 - L'^2 \right ) \cos^2 \theta}{L^2 \left ( \cos^2 \theta -1 \right )} \, .
\ee
The non-singular bulk solutions therefore have to satisfy $\gamma^2 = \beta^2$
and 
\be
L'^2 \sim \beta^2  \quad \mbox{and} \quad
A'\sim 0 \, ,
\ee
as $r$ goes to zero, which imposes the topology of a conical deficit angle. However,
for codimension 4, a jump in $A'$ does not give rise to a Dirac
distribution since the second derivatives of $A$ only appear in terms of
order $L^{-2}$ in the components of the Gauss-Bonnet tensor. We
therefore set $A \sim 1$ as $r$ goes to zero. In the vicinity of the brane, we thus obtain
\ba
H^t{}_t = H^z{}_z &\sim& \frac{4}{L^3} \hat{f}' + \frac{4\delta(r)}{L^3} \left [ f \right ]\label{themis}\\
H^r{}_r &\sim& 4 \frac{k^2 \left (L'^2- \beta^2 \right )}{L^2}  \\
H^\theta{}_\theta = H^\phi{}_\phi = H^\psi{}_\psi &\sim& \frac{4 k^2 \left
( L'^2 - \beta^2 \right )}{L^2} - 8k^2 \frac{L''}{L} \, ,
\ea
where we have set $f=(3\beta^2-L'^2)L'$. The jump of this function at the
origin, 
\be
\label{Lcone2}
f(r) = \left \{ \begin{array}{cc}
 2\beta^3 & \mbox{if $r = 0^+$} \\
3\beta^2-1 & \mbox{if $r=0$}
\end{array} \right .  
\ee
allows for the topological matching condition 
\be
\label{Codim4jc}
\sigma={2\pi^2 \over \beta^2} (1-\beta)(1+\beta-2\beta^2)\alpha_2 \, .
\ee
As in
the previous example, the Gauss-Bonnet tensor carries the charge for the
1-brane of codimension 4. Notice that (\ref{Codim4jc}) reduces to a relation between the angle
deficit of the bulk and the tension of the brane. This is very similar 
to what happens for codimension 2 branes in Einstein
gravity in the following sense: in the latter case, the Einstein bulk tensor  is reduced to a
cosmological constant term at the tip of a 2-dimensional conical
defect. Here, the Gauss-Bonnet tensor is reduced to a cosmological
constant term at the tip of a 4-dimensional defect. This result could be
derived by a direct application of Chern's theorem to
the normal section. Indeed, (\ref{themis}) is just the
Gauss-Bonnet scalar, whose integration over the normal section yields again
(\ref{Codim4jc}). Similarily, one can obtain the cosmic string tension by
application of the Gauss-Bonnet theorem on the Ricci scalar integrated over
the 2-dimensional cone \cite{vilenkin}. As for the codimension 3 example, we
point out that our solution has a conical singularity along the polar axis
$\theta=\phi=0$ if we dont want to introduce a naked singularity at $r=0$.

As in the previous
section, we extend the above result to 
Lovelock theory, including  the Einstein and
cosmological constant terms. For maximal symmetry on the bulk and on the brane,
we have the following bulk solution,
\ba
ds^2 &=& k^2 r_0^2 \, \cos^2 \left (\frac{r}{r_0} \right ) \, \left ( -dt^2 +
e^{2kt} dz^2 \right ) \nonumber \\
&& + dr^2 + r_0^2 \, \sin^2 \left (\frac{r}{r_0}
\right ) \, \left ( d\theta^2 + \sin^2 \theta \, d\phi^2 + \sin^2 \theta
\, \sin^2 \phi \, d\psi^2 \right ) \, ,
\ea
with
\be
r_0^2 = \frac{\zeta \pm \sqrt{\zeta^2 + \frac{12 \alpha
\Lambda}{5}}}{\Lambda / 5} \, .
\ee
Cutting a solid angle, by reducing the interval in which $\psi$ varies{\footnote{Note that in order not
to deform the sphere we always cut-off the defect along the Killing vector $\partial_\psi$.}},
yields an Einstein-Gauss-Bonnet 1-brane of codimension 4, embedded in a
constant curvature bulk with charge (\ref{Codim4jc}). 

It is fairly easy now (although in component language technically involved)
to extend  the maximally symmetric solutions
given here, in order to describe $p$-branes of higher dimension. If the bulk
spacetime is flat, we work out the Lovelock densities carrying charge according
to (\ref{odd}) or (\ref{fax}) and introduce an overall angular defect in the
direction of the axial Killing vector to match with
the brane tension. If bulk spacetime has constant curvature, one uses the
relevant slicing of dS (or adS) space, 
\be
ds^2 = k^2 r_0^2 \, \cos^2 \left (\frac{r}{r_0} \right ) \, \left ( -dt^2 +
e^{2kt} dz_p^2 \right ) + dr^2 + r_0^2 \, \sin^2 \left (\frac{r}{r_0}
\right ) \, d\Omega^2_{N-1} \, ,
\ee
and proceeds in a similar way as for the flat case.

\section{Summary and discussion}

We have derived matching conditions for Lovelock gravity in
arbitrary codimension. These describe the equations of
motion of a zero thickness $p$-brane, sourced by its localised
energy-momentum tensor. Self-gravity of the brane is therefore taken into
account. We have shown that unlike the situation in Einstein
gravity, defects of codimension strictly higher than 2 are possible in
Lovelock theory, as long as (\ref{special}) is satisfied. Examples of  maximally symmetric solutions were given in Section 6,
where we saw in particular that spacetime was locally regular with a
Dirac charge given by a solid angular defect.  In mathematical terms, three
types of matching conditions are possible, depending on the nature of the
geometric terms providing the charge. Mathematical regularity, along with the
examples of section 6, point towards topological
matching conditions if codimension is even (\ref{topo}) and mixed matching
conditions (\ref{matching}) if codimension is odd. 
   
In even codimensions $N=2n$, the geometric terms
appearing are, dropping out the parallel indices for simplicity, induced curvature terms $\Omega$ (\ref{led}) and even powers of the
extrinsic curvature $\k$ (\ref{ext121}): schematically $\k^2$, $\Omega_{ind}$,
$\k^4$, $\Omega^2_{ind}$, $\k^2 \Omega_{ind}$ {\it etc}, as is shown in
the table. The most striking feature
of these matching conditions is the dimensional reduction of Lovelock
densities that takes
place at the location of the brane. We see that the highest
rank Lovelock densities in the bulk $\L_{(k)}$, which carry the distributional
charge, are reduced to induced Lovelock densities on the brane $\L^{ind}_{(k-n)}$. This results in the presence
of the induced Einstein tensor in the equations of motion for $p>2$. Furthermore, if $n\geq {p+1\over 2}$, there are no
extrinsic curvature corrections and the equations of motion obeyed by the
brane are those of induced Lovelock gravity. The only degree of freedom
originating from the bulk spacetime is the topological charge $\beta(P)$,  which
along with the coupling constant $\alpha_k$ fixes the Planck scale for
4-dimensional gravity. Note that $\beta$ being a topological charge, it is not
fixed by the local field equation (\ref{ll}). Indeed, in all generality,
$\beta$ is a smooth function of the position along the brane and it is not constrained by the matching
conditions. In order to determine the dynamics --if any-- of $\beta$, one has
to solve the full Lovelock equations. Local gravity experiments put
however stringent constraints on $\beta$, which has to be varying very slowly today,
in order not to violate in any noticable way the equivalence principle
\cite{gdot}. Since the coupling of geometry to matter would vary in this case,
the variation of $\beta$ may correspond to energy loss in the extra dimensions which would
be interesting to investigate in relation to the late time acceleration of our
universe. 

As an example here, let us consider the case of a Dirac 3-brane in $D=10$
dimensions and take $\beta$ constant. The
action describing the brane's motion is (\ref{robin}), \ie the Einstein-Hilbert
plus cosmological constant. The Einstein term is obtained  from $\L_{(4)}$, the
highest order Lovelock term, while  the induced cosmological
constant comes from $\L_{(3)}$. The 10 dimensional cosmological constant,
Einstein and Gauss-Bonnet terms, $k=0,1,2$, do not carry any distributional charge and
therefore do not appear at the level of an infinitesimally thin brane. An observer on the 3-brane, if it has no access to finite thickness
corrections and $\beta$ is constant,  would never realise that it were part of a higher dimensional
spacetime. The perturbation operator for small fluctuations in the
$\mu\nu$-directions (tangent to the brane) would have to be
exactly that of ordinary 4 dimensional gravity.  Is 4 dimensional gravity
exactly localised ? Are there additional degrees of freedom hiding in the
remaining Lovelock equations and how are these perceived by a brane observer?
Clearly the dynamics of $\beta (P)$ are a key question here.
These questions certainly demand further and careful investigation. For
constant $\beta$ however, we can summarise our findings in the following way: the distributional
part originating from 
most general action in the bulk equations yields --for even codimensions-- the most general
induced action on the brane. In particular, induced gravity terms \cite{dgp} are naturally
obtained this way in a classical context. This is pictured schematically in the following
table, where, note, that under the diagonal, one only finds induced Lovelock
densities and all extrinsic curvature terms cancel out. 

We would like to remark that it is straightforward to show
that our results agree with what has been found in \cite{Navarro} (see also
the earlier work of \cite{Kaloper}) concerning
brane intersections. To recover for example the results of Navarro and
Santiago, we take a geodesic embedding of $\Sigma$ into a spacetime with topology
\be
\label{coneprod}
M=\Sigma \times \left ({\mathbb C}_2 \right )^{N/2} \, ,
\ee
where ${\mathbb C}_2$ is the 2-dimensional cone and apply the factorization rule of
the Euler density for two manifolds $M$ and $N$
\be
{\cal L} \left [ TM \oplus TN \right ] = {\cal L} \left [ TM \right ]
\wedge {\cal L} \left [ TN \right ] \, .
\ee

Also, using the action (\ref{melina}) for a codimension 2 brane
relates  the entropy of Lovelock black holes \cite{Fursaev,Carlip,Banados,Jacobson,Susskind,doug,Cai} and codimension 2 matching conditions \cite{Greg}. The
Euclidean version of (\ref{melina}), with extrinsic curvature set to zero,
yields, after using the techniques of \cite{Fursaev,Carlip,Banados,Jacobson,Susskind} 
\be
{\mathcal S}_{BH} \equiv \beta \frac{\partial
S_D^{(Eucl.)}}{\partial \beta} - S_D^{(Eucl.)}=4 \pi \int_{\Sigma} \theta^{\star \pp} \left (\alpha_1+  \alpha_2
R^{ind} \right ) \, ,
\ee
in exact agreement with \cite{doug}, which uses the Noether charge method developped in \cite{Iyer,IW}.

\vspace*{5mm}
\noindent
\renewcommand{\arraystretch}{1}
\begin{tabular}{||c|c|c|c|c|c|c||}\hline \hline
$n\backslash \tilde{k}$& $0$ & $1$ & $2$ & $3$ &
\multicolumn{2}{c||}{$\cdots$  $\left [p/2
\right ] $} \\
\hline\hline
$n=1$ & ${\mathbb{ I}}$ & $\Omega - \k^2$ & $\frac{1}{2} (\Omega^2+\k^4) - \Omega\k^2 $ &
$\frac{1}{6}(\Omega^3 - \k^6) + \frac{1}{2} (\Omega\k^4 - \Omega^2\k^2) $ &
\multicolumn{2}{c||}{
$\cdots$} \\
\cline{1-1}\cline{3-7}
$n=2$ & & $\Omega$ & $\frac{1}{2} (\Omega^2 - \k^4)$ & $\frac{1}{6}
(\Omega^3-3\Omega\k^4+2\k^6) $ & \multicolumn{2}{c||}{
$\cdots$}  \\
\cline{1-1}\cline{4-7}
$n=3$ & & & $\frac{1}{2} \Omega^2$ & $\frac{1}{6} (\Omega^3 -\Omega \k^4 ) $ & \multicolumn{2}{c||}{
$\cdots$}
 \\
\cline{1-1}\cline{5-7}
$n=4$ & & & & $\frac{1}{6} \Omega^3$ & \multicolumn{2}{c||}{
$\cdots$}   \\
\cline{1-1}\cline{6-7}
$\vdots$ & & & & & \multicolumn{2}{c||}{
$\ddots$}  \\
\hline \hline
\end{tabular}
\vspace*{5mm}

In odd codimension, we end up with Israel-like matching conditions
(\ref{matching}), 
that link the discontinuity of some quantities, to the stress-energy density
of the brane. 
These quantities determine the brane's motion  and are {\it odd} powers of $\k$, \ie
schematicaly $\k$, $\R \k$, $\k^3$ {\it etc}, as shown in the following
table. Note that if we consider axial symmetry, the terms beneath the diagonal
are dimensionally extended Chern-Simons forms, \ie the usual boundary terms
(\ref{zztop}), (\ref{my}). This is typical of the 
motion of a singular hypersurface, where in the context of branewords it 
is known, for example,  that only at late times is the cosmological evolution
4 dimensional \cite{bin2}. This is a clean cut difference from even
codimension matching conditions.

\vspace*{5mm}
\noindent
\renewcommand{\arraystretch}{1}
\begin{tabular}{||c|c|c|c|c|c||}\hline \hline
$n\backslash \tilde{k}$ & $1$ & $2$ & $3$ & \multicolumn{2}{c||}{$\cdots$
  $\left [(p+1)/2 \right ]$} \\
\hline\hline
$n=1$ & $\k$ & $\R\k$ & $\frac{1}{2}\R^2\k$ & \multicolumn{2}{c||}{$\cdots$}  \\
\cline{1-1}\cline{3-6}
$n=2$ & & $\R\k+ \frac{2}{3} \k^3 $ & $\frac{1}{2} \R^2\k+\frac{2}{3}\R\k^3$ &  \multicolumn{2}{c||}{$\cdots$}  \\
\cline{1-1}\cline{4-6}
$n=3$ & & & $\frac{1}{2} \R^2 \k + \frac{2}{3}\R\k^3+\frac{4}{15}\k^5$ & \multicolumn{2}{c||}{$\cdots$}  \\
\cline{1-1}\cline{5-6}
$\vdots$ & & &  & \multicolumn{2}{c||}{$\ddots$} \\
\hline \hline
\end{tabular}
\vspace*{5mm}

Furthermore, and this is an important difference from hypersurfaces, as the
codimension increases, the brane's motion is less and less constrained by the
bulk. This we saw in particular for the 3-brane's motion in codimension 2,
where extrinsic curvature corrections were present, only to dissapear for
$n>2$. {\it The matching equations are necessary but not sufficient
conditions}. What we have obtained here is the purely distributional part of the Lovelock bulk
equations, which yields quite naturally, sensible matching conditions for the
distributional p-brane.
 It can be that a singular behavior of the Riemann tensor, in the
vicinity of the brane, makes these unfeasible for a concrete exact solution. 
Actually, one can expect this from classical black hole theorems. 
We know that Birkhoff's theorem is true for Einstein-Gauss-Bonnet gravity and,
since the 
degrees of freedom for the graviton do not change
in Lovelock theory{\footnote{We thank Roberto Emparan for pointing this
    out.}},  we expect the theorem to hold there too. Therefore, if the bulk
solution is not locally flat, as for the simple examples we have described,
then, we expect the formation of horizons, or worse, of naked singularities in
the vicinity of the brane. The analogue solutions have been analysed by Gregory
\cite{ruth} and the braneworld picture has been discussed in
\cite{emp}. It may be that constraints on the curvature
may make the junction conditions unfeasible once gravitons freely propagate
in the bulk. This would simply mean that a Dirac distribution is then too crude
an approximation to describe in a consistent way the brane's motion. 
This is also matter for future investigation.

\section*{Acknowledgements}
\noindent
It is a pleasure to thank Martin Bucher, Brandon Carter, Roberto Emparan, Ruth
Gregory, 
Tony Padilla from Liverpool, Jos\'e Santiago and in particular
Dani\`ele Steer for many interesting discussions, comments and 
encouraging remarks. We would also particularily like to thank Giorgos Kofinas for
discussions on energy conservation and the referee for constructive
comments and remarks on our initial manuscript concerning in particular
sections 4, 5  and 6.

\appendix
\section{Definitions} 
\label{op}

\subsection{The Dirac distribution}
The first thing we need to define is the Dirac distribution centered on a
submanifold $\Sigma$ of the total spacetime $M$, so that we can account for
the localisation of stress-energy over $\Sigma$. To do so, let $U$ be an arbitrary open submanifold of $M$ such that
\be
\mbox{Codim}_U (U \cap \Sigma ) = N \, ,
\ee
\ie $U$ covers all the normal directions to $\Sigma$. Then, to any function $f:U \longrightarrow {\mathbb R}$, of compact support in
$U$, the Dirac distribution centered on $\Sigma$, $\delta_{\Sigma}$,
associates, by definition, the real value 
\be
\int_U f\delta_{\Sigma} = \left\{  \begin{array}{cc}
 f(P) & \mbox{if $U \cap \Sigma = \{ P \}$} \\
 \int_{U\cap \Sigma} f & \mbox{otherwise.}
\end{array} \right .
\ee

\subsection{Operators on $\Omega^*(TM)$}
In this appendix, we also define some useful operators over the space of
differential forms $\Omega^*(TM)$, that acts on the tangent bundle of
$(M,g)$. We first consider projection operators. These are built out of
the interior product. The latter is defined over $\Omega^{(r)}(TM)$ by,
\ba
\label{int}
i : TM \times \Omega^{(r)}(TM) &\rightarrow & \Omega^{(r-1)}(TM) \nonumber \\
(X,\omega) &\rightarrow & i_X \omega \, ,
\ea
with 
\be
(i_X \omega) (X_1,\cdots,X_{r-1})=\omega (X,X_1,\cdots,X_{r-1}) \, ,
\ee
for all $X_1$, $\cdots$, $X_{r-1}$ in $TM$. We then define the projection operator from $\Omega^{(r)}(TM)$ to
$\Omega^{(r-q)}(T\Sigma) \otimes \Omega^{(q)}(T\Sigma^\perp)$ by
\be
\label{int2}
i_{q} \equiv  
\left ( \bigwedge_{l=1}^{r-q} \theta^{\mu_l} \right ) \wedge \left (\bigwedge_{l=1}^q \theta^{I_l} \right ) i_{n_{I_q}}
\cdots i_{n_{I_1}} \left(i_{e_{\mu_{r-q}}} \cdots i_{e_{\mu_{1}}}\right)
\ee
The special case $r=D-1$, $q=N$ is particularly helpful in this paper. The
corresponding projection operator
\be
\label{projector}
i_{N} \equiv 
\left ( \bigwedge_{l=1}^{D-N-1} \theta^{\mu_l} \right ) \wedge \left
(\bigwedge_{l=1}^N \theta^{I_l} \right ) i_{n_{I_N}} \cdots i_{n_{I_1}} \left(i_{e_{\mu_{D-N-1}}} \cdots i_{e_{\mu_{1}}}\right)
\ee 
achieves two things. The $i_{e_\mu}$ terms in parentheses  project
all tangent directions but one, factorizing an $N$-form from the total
$(D-1)$-form and the $i_{n_I}$ terms project this $N$-form uniquely on the
normal section $(T\Sigma)^{\perp}$. This is precisely what is needed in
section \ref{gauss}, in order to integrate (\ref{match0}). Developping every
$(D-1)$-form (\ref{klov}) according to the decomposition of the curvature
2-form (\ref{proj}), we get a sum over all the relevant powers of $\R_{\pp}$,
$\tilde{\R}$ and $\R_{\perp}$. The latter, once projected by means of $i_N$,
reads 
\ba
\label{ein}
i_N{\cal E}_{(k) A} &=& \sum_{i=0}^{k} \sum_{j=0}^{k-i} C_k^i
C_{k-i}^j \left ( \bigwedge_{l=1}^j \R^{A_l B_l}_{{\pp}} \right )\left (\bigwedge_{l=1}^{k-i-j} \tilde{\R}^{C_l
D_l}{} \right ) \left (\bigwedge_{l=1}^{i} \R^{E_l F_l}_{\perp}
\right ) \nonumber \\
&& \qquad \qquad \qquad \qquad \qquad \qquad \qquad \wedge \, i_{N-k-i+j}\theta^{\star}_{AA_1\cdots B_jC_1 \cdots
D_{k-i-j}E_1 \cdots F_i} \, ,
\ea
where as usual, $C^i_j={j!\over i! (j-i)!}$. Since the exterior product of the
curvature 2-form projections is a $2k$-form with tensor structure
$\Omega^{(k-i+j)}(T\Sigma)\otimes\Omega^{(k+i-j)}(T\Sigma^\perp)$, 
the remaining $i_{N-k-i+j}\theta^\star_{A \cdots}$ are $(D-2k-1)$-forms, with
tensor structure $\Omega^{(D-N-k+i-j-1)}(T\Sigma) \otimes
\Omega^{(N-k-i+j)}(T\Sigma^\perp)$. This makes $i_N{\cal E}_{(k) A}$ an $N$-form on $(T\Sigma)^{\perp}$ as was required.

With respect to these projections, one can also split the differential
operator $d$ into parallel and perpendicular differential operators,
respectively $d_{{\pp}}$ and $d_\perp$. These act as projected operators on
$\Omega^*(T\Sigma)$ and $\Omega^*(T\Sigma^\perp)$. To summarize, we have the following diagram
$$\xymatrix{ \Omega^{(r-q)}(T\Sigma) \otimes
    \Omega^{(q)}(T\Sigma^\perp) \ar[rr]^-{d_\perp} \ar[dd]^-{d_{{\pp}}} && \Omega^{(r-q)}(T\Sigma) \otimes
    \Omega^{(q+1)}(T\Sigma^\perp) \ar[dd]^-{d_{{\pp}}}  \\
&  \Omega^{(r+1)}(TM) \ar[ru]^-{i_{q+1}} \ar[ld]^-{i_q} &\\
 \Omega^{(r-q+1)}(T\Sigma) \otimes
    \Omega^{(q)}(T\Sigma^\perp) \ar[rr]^-{d_\perp}  && \Omega^{(r-q+1)}(T\Sigma) \otimes
    \Omega^{(q+1)}(T\Sigma^\perp)  } $$

However, it is worth emphasizing that $d_\perp$ {\it is not} the proper
differential operator $d_{\Sigma_P^\perp}$ of $\Omega^*(T\Sigma^\perp)$,
but only the projection onto it of the spacetime differential operator
$d$. Indeed, one has 
\be
d_\perp = (-1)^{r}  d_{\Sigma_P^\perp} \, ,
\ee 
over $\Omega^{(r)}(T\Sigma) \otimes \Omega^{(q)}(T\Sigma^\perp)$. The
differential forms of $\Omega^*(T\Sigma)$ are thus seen as such by $d_\perp$,
while they are seen as mere functions by $d_{\Sigma_P^\perp}$. A further
consequence is the appearance of an overall sign when applying Stoke's theorem to $d_\perp$, 
\be
\int_{\Sigma_P^\perp} d_{\perp} \omega = (-1)^{r} \int_{\Sigma_P^\perp}
d_{\Sigma_P^\perp} \omega =  (-1)^{r} \int_{\partial \Sigma_P^\perp} \omega
\, ,
\ee 
for all $\omega$ in $\Omega^{(r)}(T\Sigma) \otimes \Omega^{(N-1)}(T\Sigma^\perp)$.

\section{Normal connection-invariant cohomology classes}
\label{cohom}
Consider an arbitrary $(D-1)$-form $\Xi_{(k)A}$ from (\ref{ein}). We suppose
that it depends explicitely on the normal connection $\psi^{IJ}_\perp$ and not
only on the local normal frame $\theta^I$.  
We shall now demonstrate that the integral of $\Xi_{(k)A}$ over $\Sigma_P^\perp$
is invariant under (\ref{coc}), if and only if it is locally exact in the
sense of $d_\perp$. By locally exact, we mean that its pull-back to the
associated sphere bundle $\pi_0: {\mathbb S}_{N-1} \rightarrow
\Sigma_P^\perp\backslash \{ P \}$ is exact. 


So, let $\Xi_{(k)A}$ be some term in (\ref{ein}). If $\pi^*_0 (\Xi_{(k)A})$ is exact, there exists some $\sigma_{(k)A}$ such that 
\be
\pi^*_0 (\Xi_{(k)A}) = d_\perp\sigma_{(k)A} \, .
\ee
We thus have,
\be
\int_{\Sigma_P^\perp} \Xi_{(k)A} = \int_{n_N(\partial (\Sigma_P^\perp \backslash
  \{ P\}))} \sigma_{(k)A} \, .
\ee
Varying with respect to $t$, one gets,
\be
\frac{d}{dt} \int_{\Sigma_P^\perp} \Xi_{(k)A} = \int_{n_N(\partial
  (\Sigma_P^\perp \backslash  \{ P\}))}  \xi^{IJ} \wedge (\cdots) 
= 0 \,  ,
\ee
since $\xi^{IJ}$ vanishes on the boundary. We thus show that the integral over
$\Sigma_P^\perp$ of the locally exact forms are normal connection-invariant.

On the other hand, if $\Xi_{(k)A}$ is not locally exact, one has 
\be
\label{isnotexact}
\frac{d}{dt} \Xi_{(k)A)} = \xi^{IJ} \wedge \lambda_{(k)AIJ}(t) \, .
\ee
Notice that one could in principle find terms of the form
\be
\frac{d}{dt} \Xi_{(k)A)} =( d_\perp \xi^{IJ}) \wedge B_{(k)AIJ}(t) \, ,
\ee
but these terms can be integrated by parts according to 
\be
\left (d_\perp \xi^{IJ} \right ) \wedge B_{(k)AIJ}(t) = d_\perp \left
  (\xi^{IJ}\wedge B_{(k)AIJ}(t) \right ) + \xi^{IJ} \wedge \left (d_\perp
  B_{(k)AIJ}(t) \right ) \, , 
\ee
yielding either locally exact terms or terms like (\ref{isnotexact}).
Then,
\be
\frac{d}{dt} \int_{\Sigma_P^\perp} \Xi_{(k)A} = \int_{\Sigma_P^\perp}
\frac{d\Xi_{(k)A} }{dt} = \int_{\Sigma_P^\perp} \xi^{IJ} \wedge
\lambda_{(k)AIJ}(t) \, .
\ee
We therefore have normal connection invariant integrals for those terms, if
and only if,
\be
\int_{\Sigma_P^\perp} \xi^{IJ} \wedge
\lambda_{(k)AIJ}(t) = 0 \, ,
\ee
for all $t$ in $[0,1]$ and all $\xi^{IJ}$. This
implies that for all $t$ in $[0,1]$, $\lambda_{(k)AIJ}(t) = 0$ over
$\Sigma_P^\perp$ and then, after (\ref{isnotexact}), that $\Xi_{(k)A}$ is
independent of $t$. The latter point can only be achieved for terms that do
not depend on the normal connection at all, which contradicts our hypothesis. The only terms in (\ref{ein})
explicitely depending on the normal connection and whose integrals over
$\Sigma_P^\perp$ are nonetheless normal connection-invariant are therefore the
locally exact ones. Notice that requiring an explicit dependance on the normal
connection for $\Xi_{(k)A}$ is necessary in order for it to be charged, \ie to have a non-vanishing integral over $\Sigma_P^\perp$ even when the
latter is shrunk to $P$.

We thus see that when integrating (\ref{ein}) over $\Sigma_P^\perp$, the only
normal connection-invariant integrals of charged terms come from the terms in
(\ref{ein}) that pull back to the null cohomology class over the sphere bundle
$\left [ 0 \right ] \in H^{(N)}({\mathbb S}_{N-1})$. This result might seem
trivial, but one has to think that when pulled back to the tangent bundle
$T\Sigma^\perp \rightarrow \Sigma_P^\perp$ of the normal section, it yields
non-trivial cohomology classes of $H^{(N)}(T\Sigma^\perp)$. Indeed, it
contains for example the Euler cohomology class $\left [{\mathcal
    L}^\perp_{(N/2)} \right ] \in H^{(N)}(T\Sigma^\perp)$ which plays a
central role in the form of the matching conditions in even
codimensions.~\footnote{A pull back to the principal fiber bundle $\pi:
  O(D-N)\rightarrow \Sigma_P^\perp$ also yields its Pontrjagin characteristic class.}

\section{Charged exact forms}
\label{appb}

Since the normal frame vanishes at $P$, the basis of dual 1-forms $\theta^I$ identically pulls back to the null 1-form over the sphere bundle. We thus
conclude that the charged exact forms can not contain any such
form. This has two major consequences. The first one is that the charged
exact forms are contained in the terms of (\ref{ein}) for which
\be
\label{itheta}
i_{N-k-i+j} \theta^\star_{\mu \cdots} = i_0 \theta^\star_{\mu \cdots}\, ,
\ee
which leads to $k-N+i-j=0$. The second is that the charged exact forms
must contain the maximal power of the $\psi^I_{\perp N}$, which, being
the only non-vanishing normal forms at $P$, will pull back to
non-vanishing forms over the sphere bundle. With regards to this, it
will appear particularly helpful, as we shall see, to introduce the
Chern-Simons $N-1$ and $N$-forms,
\ba
\Pi_{(N,m)}^\perp &\equiv &\epsilon_{\dot{I}_1 \cdots
\dot{I}_{N-1}} \left (\bigwedge_{l=1}^{m} \Omega_{\perp}^{\dot{I}_{2l-1}
\dot{I}_{2l}} \right ) \wedge \left (\bigwedge_{l=2m+1}^{N-1}
\psi^{\dot{I}_l}_{\perp N} \right ) \, , \\ 
\Theta^\perp_{(N,m)} &\equiv & 2(m+1) \epsilon_{\dot{I}_1 \cdots \dot{I}_{N-1}} \left
(\bigwedge_{l=1}^{m} \Omega^{\dot{I}_{2l-1} \dot{I}_{2l}}_\perp \right
) \wedge \Omega^{\dot{I}_{2m+1} N}_\perp
\wedge \left (\bigwedge_{l=2m+2}^{N-1} \psi^{\dot{I}_{l}}_{\perp N}
\right ) \, ,
\ea
since the potentials and charged exact forms can be
decomposed upon both sets. Namely, taking (\ref{itheta}) into account,
we find that the charged exact forms can be written, up to regular terms, as
\be
\label{cef1}
\Xi_{(N,k)\mu} \sim \sum_{i=\mbox{\scriptsize{max}}(0,N-k)}^{\mbox{\scriptsize{min}}(k,\left [N/2 \right ])} C_k^i C_{k-i}^{k-N+i} \left
(\bigwedge_{l=1}^{k-N+i}\R_{{\pp}}^{A_l B_l} \right ) \wedge \left
( \bigwedge_{l=1}^{N-2i}\tilde{\R}^{C_l D_l} \right ) \wedge \left
( \bigwedge_{l=1}^{i} \R_{{\perp}}^{E_l F_l} \right ) \wedge i_0
\theta^\star_{\mu \cdots} \, .
\ee
Notice that (\ref{itheta}) gives a further constraint since $k-N=j-i$
for $0 \leq j \leq k-i$ and $0 \leq i \leq k$ implies that $2k \geq
N$. The sum in (\ref{cef1}) thus runs from $\mbox{max}(0,N-k)$ to $\left
[ N/2 \right ]$. Now, one needs two things. First, according to appendix \ref{cohom}, the exact forms are derived by integration by parts of some
differential $d_\perp$, meaning that we can restrict ourselves to the terms of
(\ref{cef1}) that contain a $d_\perp$. Next, these terms must contain the
maximal power of the $\psi^I_{\perp N}$ since the latter are the only
non-vanishing 1-forms at $P$. Looking at the Gauss-Codazzi equations
(\ref{gcmunu}), (\ref{gcmui}) and (\ref{gcij}), one sees that the charged
exact forms are therefore deduced from (\ref{cef1}) by trading each
$\tilde{R}^{C_l D_l}$ for 
$$D_\perp \k^{\mu_l I_l}_{{\pp}} = d_\perp \k^{\mu_l I_l}_{{\pp}} + \psi^{\mu_l}_{\perp\nu} \wedge
\k^{\nu I_l}_{{\pp}} + \k^{\mu_l}_{{\pp}J} \wedge \psi^{J I_l}_\perp $$
and each $R_{{\perp}}^{E_l F_l}$ for $\Omega_\perp^{I_l J_l}$. This yields
\be
\label{cef2}
\Xi_{(N,k)\mu} \sim \sum_{i=\mbox{\scriptsize{max}}(0,N-k)}^{\left [N/2 \right ]} 2^{N-2i}
\, C_k^i C_{k-i}^{k-N+i} \left (\bigwedge_{l=1}^{k-N+i}
\R_{{\pp}}^{\mu_l \nu_l} \right ) \wedge \left ( \bigwedge_{l=1}^{N-2i} D_\perp \k^{\sigma_l K_l}_{{\pp}} \right )
\wedge \left ( \bigwedge_{l=1}^{i} \Omega_{{\perp}}^{I_l J_l} \right )
\wedge i_0 \theta^\star_{\mu \cdots} \, .
\ee
Using 
\be
i_0 \theta^\star_{\mu \mu_1 \cdots \mu_{D-2k-2} I_1 \cdots I_N} = (-1)^{(D-2k-1)N}
\theta^{\star \pp}_{\mu \mu_1 \cdots \mu_{D-2k-2}} \epsilon_{I_1 \cdots I_N} \, ,
\ee
one can rearrange (\ref{cef2}) to get
\ba
\label{cef3}
&&\Xi_{(N,k)\mu} \sim \sum_{i=\mbox{\scriptsize{max}}(0,N-k)}^{\left [N/2 \right ]} 2^{N-2i} \, C_k^i C_{k-i}^{k-N+i} \left
(\bigwedge_{l=1}^{k-N+i} \R_{{\pp}}^{\mu_l \nu_l} \right ) \nonumber \\
&& \qquad \qquad \qquad \qquad \qquad \qquad \wedge \left [ (-1)^{D-N}(N-2i) \, d_\perp \k^{\sigma_1}_{{\pp} N} \wedge \left
( \bigwedge_{l=2}^{N-2i} \k^{\sigma_l}_{{\pp} N} \right ) \wedge
\theta^{\star {\pp}}_{\mu \cdots} \wedge \Pi^\perp_{(N,i)} \right
. \nonumber \\
&& \qquad \qquad \qquad \qquad \qquad \qquad
\qquad \qquad \qquad \left . + \left ( \bigwedge_{l=1}^{N-2i}
\k^{\sigma_l}_{{\pp}N} \right ) \wedge \theta^{\star {\pp}}_{\mu \cdots} \wedge \Theta^\perp_{(N,i-1)} \right ]
\, ,
\ea
where we set $\Theta^\perp_{(p,-1)}=\Theta^\perp_{(p,\left [ p/2 \right
])}=0$. Now, one can show \cite{Chern}, that 
\be
\label{cha}
\Theta^\perp_{(N,i)} = d_\perp \Phi^\perp_{(N,i)} \, ,
\ee
where $\Phi^\perp_{(N,i)}$ is the $(N-1)$-form defined at (\ref{z}). This, together with the identity 
\be
d_\perp \Pi^\perp_{(N,\left [N/2 \right ])} = - \Theta^\perp_{(N,\left [N/2
  \right ]-1)} = -d_\perp \Phi^\perp_{(N,\left [N/2
  \right ]-1)} \, ,
\ee
for odd values of $N$, helps in showing that 
\be
\Xi_{(N,k) \mu} = d_\perp \sigma_{(N,k) \mu} \, ,
\ee
with
\ba
\label{cef4}
\sigma_{(N,k) \mu} &=& \sum_{i=\mbox{\footnotesize max}(0,N-k)}^{\left [N/2
  \right ]} (-1)^{D-N-1} \, 2^{N-2i} C^i_k C^{k-N+i}_{k-i}
\left (\bigwedge_{l=1}^{k-N+i} \R^{\mu_{l} \nu_{l}}_{{\pp}} \right ) \wedge
\left ( \bigwedge_{l=1}^{N-2i} \k^{\rho_l}_{{\pp} N} \right ) \wedge
\theta^{\star {\pp}}_{\mu \cdots} \wedge \Phi^\perp_{(N,i-1)} \nonumber \\
&& + (-1)^{D-N} \, 2^{N-2\left [ N/2 \right ]} C_k^{\left [ N/2 \right ]}
  C_{k-\left [ N/2 \right ]}^{k-N+\left [ N/2 \right ]} \left (N-2\left [ N/2
  \right ] \right ) H_{(N,k)\mu} \, .
\ea
In the odd codimension case, one has to introduce a special correction, which
  is the second term in the above expression. In this term, one has
\ba
\label{z1}
H_{(N,k)\mu} &=& \sum_{i=0}^{k-N+\left [ N/2
  \right ]} \frac{(-1)^{k-N+\left [ N/2
  \right ]-i}}{2k-N-2i} C_{k-N+\left [ N/2
  \right ]}^i \left ( \bigwedge_{l=1}^i \underline{\R}^{\mu_l \nu_l}_{{\pp}}\right ) \wedge \left (
  \bigwedge_{l=1}^{2k-N-2i} \k^{\sigma_l}_{{\pp}N}\right ) \wedge \theta^{\star {\pp}}_{\mu
  \cdots} \nonumber \\
&& \wedge \left (\Pi^\perp_{(N,\left [ N/2
  \right ])} + \Phi^\perp_{(N,\left [ N/2
  \right ] -1)} \right ) \, ,
\ea
where we have decomposed the parallel projection of the curvature 2-form
  according to
\be
\R^{\mu \nu}_{{\pp}} = \underline{\R}^{\mu \nu}_{{\pp}} - \k^\mu_{{\pp} N}
  \wedge \k^\nu_{{\pp} N} \, ,
\ee
the $\underline{\R}^{\mu \nu}_{{\pp}}$ being smooth at P. To make contact with the codimension 1 case, which can be straightforwardly
derived from (\ref{cef3}), we shall take the convention $\Phi^\perp_{(N,-1)} =
  0$ and $\Pi^\perp_{(1,0)} = 1$.
Notice that the parallel forms involved in (\ref{cef4}) can be
seen as a dimensional continuation of the Chern-Simons forms
(\ref{C-Sforms}), in the same way as each Lanczos-Lovelock theory is a
dimensional conitnuation of some Euler density. In the end, let us emphasize
  the remarkable fact that the only possible contributions to the potential
  arrange together so as to yield an exact form, the remaining terms being
  necessarily regular.

\end{document}